\documentclass{ieeeaccess}
\usepackage{cite}
\usepackage{amsmath,amssymb,amsfonts,bm}
\usepackage{algorithmic}
\usepackage{graphicx}
\usepackage{textcomp}
\usepackage{color}
\usepackage{ulem}
\usepackage{caption}
\usepackage{multirow}

\newcommand{\vct}[1]    {\mbox{\boldmath{$#1$}}}
\newcommand{\jj}        {{\rm j}}
\newcommand{\dd}        {{\rm d}}

\begin{document}
\history{Date of publication xxxx 00, 0000, date of current version xxxx 00, 0000.}
\doi{10.1109/ACCESS.2017.DOI}

\title{Noncontact Respiratory Measurement for Multiple People at Arbitrary Locations Using Array Radar and Respiratory-space Clustering}
\author{Takato~Koda\authorrefmark{1}, Takuya~Sakamoto \authorrefmark{1}~\IEEEmembership{Senior Member,~IEEE}, Shigeaki Okumura~\authorrefmark{2}, and Hirofumi~Taki~\authorrefmark{2}~\IEEEmembership{Member,~IEEE}}
\address[1]{Department of Electrical Engineering, Graduate School of Engineering, Kyoto University, Kyoto, Kyoto, 615-8510 Japan}
\address[2]{MaRI Co., Ltd., Kyoto, Kyoto, 600-8815, Japan}

\tfootnote{This study was supported in part by JSPS KAKENHI 19H02155, JST PRESTO JPMJPR1873, and JST COI JPMJCE1307.}

\markboth{}
{Koda \emph{et al.}: Noncontact Respiratory Measurement for Multiple People Using Array Radar}

\corresp{Corresponding author: Takuya~Sakamoto (e-mail: sakamoto.takuya.8n@kyoto-u.ac.jp)}

\begin{abstract}
    We developed a noncontact measurement system for monitoring the respiration of multiple people using millimeter-wave array radar. To separate the radar echoes of multiple people, conventional techniques cluster the radar echoes in the time, frequency, or spatial domain. Focusing on the measurement of the respiratory signals of multiple people, we propose a method called respiratory-space clustering, in which individual differences in the respiratory rate are effectively exploited to accurately resolve the echoes from human bodies. The proposed respiratory-space clustering can separate echoes, even when people are located close to each other. In addition, the proposed method can be applied when the number of targets is unknown and can accurately estimate the number and positions of people. We perform multiple experiments involving five or seven participants to verify the performance of the proposed method, and quantitatively evaluate the estimation accuracy for the number of people and the respiratory intervals. The experimental results show that the average root-mean-square error in estimating the respiratory interval is 196 ms using the proposed method. The use of the proposed method, rather the conventional method, improves the accuracy of the estimation of the number of people by 85.0\%, which indicates the effectiveness of the proposed method for the measurement of the respiration of multiple people. 
\end{abstract}


\titlepgskip=-15pt

\maketitle
   
\section{Introduction}
According to the World Health Organization\cite{WHO}, severe pneumonia is associated with a respiratory rate of more than 30 breaths per min. The respiratory rate is extensively used in the triage, diagnosis, and prognosis of the novel coronavirus infection, and the importance of the noncontact monitoring of respiratory rate data has been noted\cite{Massaroni}. The adoption of radar technology is a suitable approach for monitoring respiratory diseases because microwaves and millimeter-waves can penetrate clothes, bedding, and similar fabric obstacles, allowing radar systems to measure skin displacements generated by the physiological signals of distant people accurately, without requiring sensors. In this study, we developed a noncontact measurement system for respiration monitoring using a millimeter-wave array radar system based on a novel clustering algorithm that exploits individual respiratory differences, with the aim of monitoring the respiration of several people simultaneously and with high accuracy. The proposed method implements a series of processes, including the automatic detection of people, estimation of the number and locations of people, signal separation, and respiratory measurement. We thereby demonstrate accurate and noncontact respiratory measurement of several people, which cannot be realized using conventional methods.

Radar-based noncontact respiratory measurements covering multiple human targets have been extensively studied\cite{3,4,5,6,7,8,9,add1,add2,add3,10,11,12,13,14,15,16,add4,add5}. However, such conventional studies have targeted only two participants\cite{3,4,5,6,7,8,9,add1,add2,add3} or a handful of participants\cite{10,11,12,13,14,15,16,add4,add5}. Most previous studies on the noncontact respiratory measurement of several people assumed ideal conditions; e.g., Yang et al.\cite{11} and Su et al.\cite{14} assumed a special case where all participants were at different distances from the radar antenna while Xiong et al.\cite{10} and Ding et al.\cite{12} assumed a special case where all participants were lined up in a straight line. Excluding studies that impose such special conditions, there are no examples of simultaneous noncontact respiratory measurement involving several people to the best of our knowledge.

Some previous studies focused only on the detection of respiration\cite{7,16,17} whereas others demonstrated the estimation of the respiration rate. In addition, respiratory rates obtained using a reference respiratory monitor belt and a radar system have been compared\cite{5,9,10,11,12}. For example, Lu et al.\cite{9} quantitatively evaluated the respiratory-rate estimation accuracy for two participants. However, no study has quantitatively evaluated the accuracy of respiratory rate estimation for several people using radar. In this study, we quantitatively evaluate the accuracy of the estimated respiratory rate of several people.

Furthermore, most previous studies assumed that the number of people to be measured is known in advance. However, to apply such methods in a real environment, it is necessary to find a technique for estimating the number of people. Besides respiratory measurement studies, methods for estimating the number of people were proposed by Choi et al.\cite{18} and Yavari et al.\cite{19}. However, as these methods include neither target location estimation nor respiratory measurement, they are unsuitable for the purpose of this study. A method for estimating the number of people by detecting their respiration was proposed by Lv et al.\cite{17}, and methods for estimating the number and locations of people by detecting their respiration were proposed by Ha et al.\cite{7} and Nov\'{a}k et al.\cite{16}; however, these methods were not designed for the measurement of the respiratory rate.

In this study, we develop a radar-based system for noncontact respiratory measurement involving several people, when the number and locations of people are unknown. More specifically, to separate multiple radar echoes, instead of the traditional approach involving the application of clustering in the real space, the proposed method applies clustering to high-dimensional space constructed by converting the estimated respiratory intervals into virtual distances used with real-space variables. Hereafter, we refer to this multidimensional space as the respiratory-space. Clustering in the respiratory-space allows for the clear separation and discrimination of multiple radar echoes from targets in proximity, and the accurate measurement of the respiration rates of several people simultaneously, even under crowded conditions. We performed radar measurement experiments with up to seven participants, and quantitatively evaluated the performance of the proposed method by comparing the results with reference data obtained from belt-type respirometer sensors worn on the chests of the participants. The experimental results demonstrate that our proposed respiratory-space clustering outperforms the conventional real-space clustering approach, realizing accurate and simultaneous measurement of the respiration of several people.

    Regarding the selection of the operating frequency, low-frequency microwaves (e.g., microwaves having a frequency of 2.4 GHz) have been used \cite{6,7,8,10,12} because devices operating at these frequencies are inexpensive and the microwaves penetrate clothing well. Radar at higher frequencies is also used because it is sensitive to small body displacements; for example, Islam et al. \cite{5}, Muragaki et al. \cite{4}, and Walterscheid et al. \cite{13} respectively used 24-, 60-, and 72-GHz radar systems for respiratory measurement. In the present study, we used a 79-GHz radar system that is suitable for detecting small body movement generated by respiration.
    As for the bandwidth, existing studies reported radar-based respiratory measurements made using signals with various bandwidths, such as 0.4 GHz \cite{3}, 1.0 GHz \cite{13,14}, 1.3 GHz \cite{4}, 1.5 GHz \cite{7}, 2.5 GHz \cite{11}, 3.0 GHz \cite{6}, 3.3 GHz \cite{15}, 4.5 GHz \cite{16}, and 8.7 GHz \cite{12}. To separate echoes from multiple people effectively, we used a radar system that had a bandwidth of 3.5 GHz in this study.

\section{Respiratory Measurement for Multiple People Using an Array Radar}
\subsection{Millimeter-wave Array Radar}
\begin{figure}[tb]
  \centering
  \includegraphics[width=0.25\textwidth]{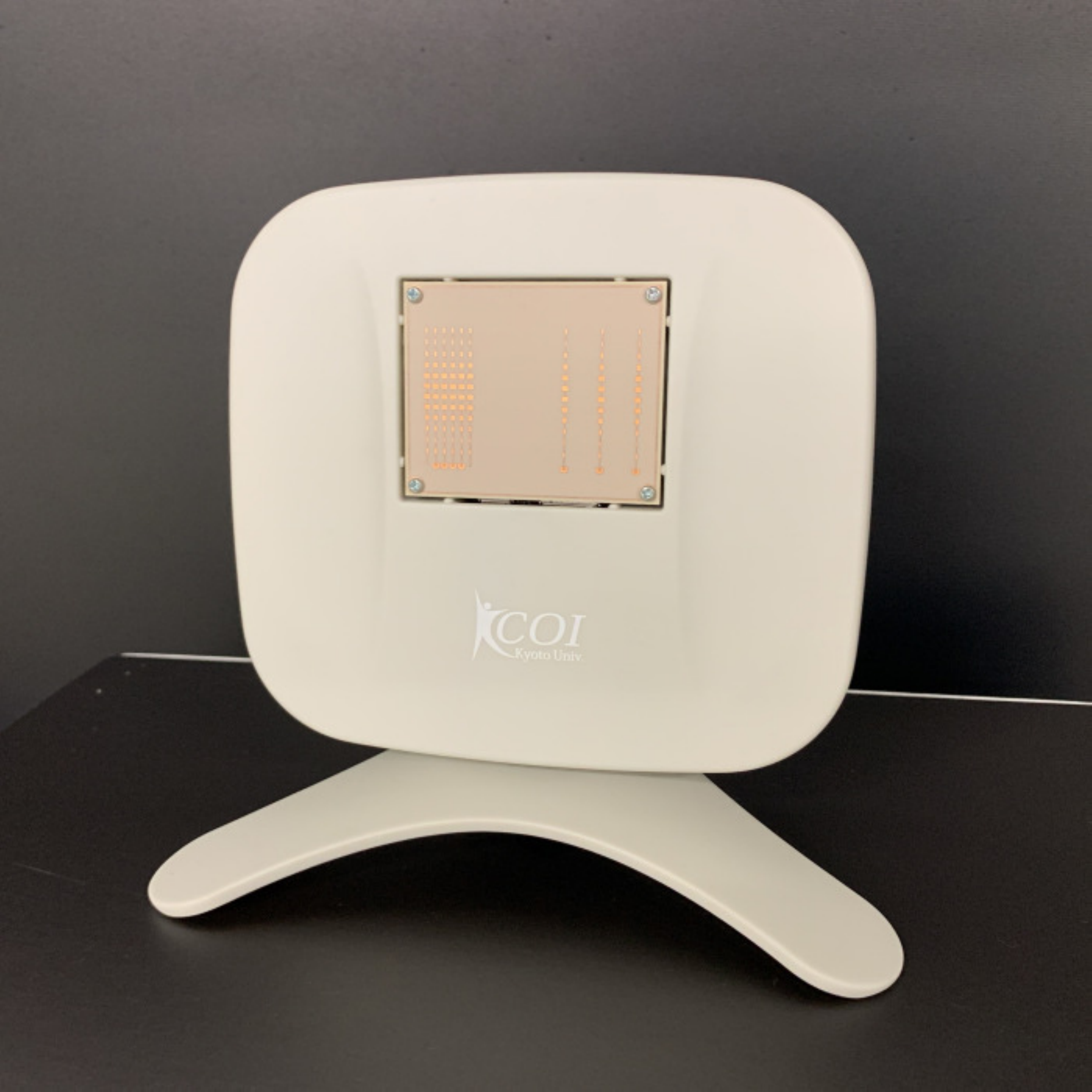}
  \caption{Photograph of the radar system developed and used in this study.}
  \label{fig:COIradar}
\end{figure}

In this study, we used a millimeter-wave multiple-input and multiple-output (MIMO) array radar system to measure the respiration of multiple people. The proposed system estimates the number and locations of the target people using the received signals and then outputs the respiratory movements and respiratory rate of each person. Figure \ref{fig:COIradar} presents a photograph of the radar system used in this study. 

This study adopted a frequency-modulated continuous-wave (FMCW) radar system with a center frequency of 79 GHz, center wavelength $\lambda$ = 3.8 mm, and a range resolution of 43 mm. The transmission power was 9 dBm, and the equivalent isotropic radiation power was 20 dBm. The system had an MIMO antenna array with three elements for transmission and four elements for reception, allowing a total of 12 signal channels to be acquired. The transmitting and receiving elements were arranged in equally spaced linear arrays at intervals of $2\lambda$ (7.6 mm) and $\lambda/2$ (1.9 mm), respectively. In general, MIMO arrays can be approximated by virtual arrays in cases where the distance between the target and antennas is sufficiently larger than the array aperture length, and the mutual coupling between the elements is negligible. The array of the above radar system can be approximated by a 12-element linear virtual array with half-wavelength spacing. The element patterns of each element were $\pm 4^{\circ} $ and $\pm 35^{\circ} $ on the E- and H-planes, respectively. 
The three transmitting elements repeated transmission with time-division multiplexing, and a total of 12 channels of the signal were stored every 100 ms, resulting in a slow-time sampling frequency of 10 Hz that is sufficiently high for respiratory measurement but too low for the measurement of the heartbeat.

\begin{figure}[tb]
    \centering
    \includegraphics[width=0.45\textwidth]{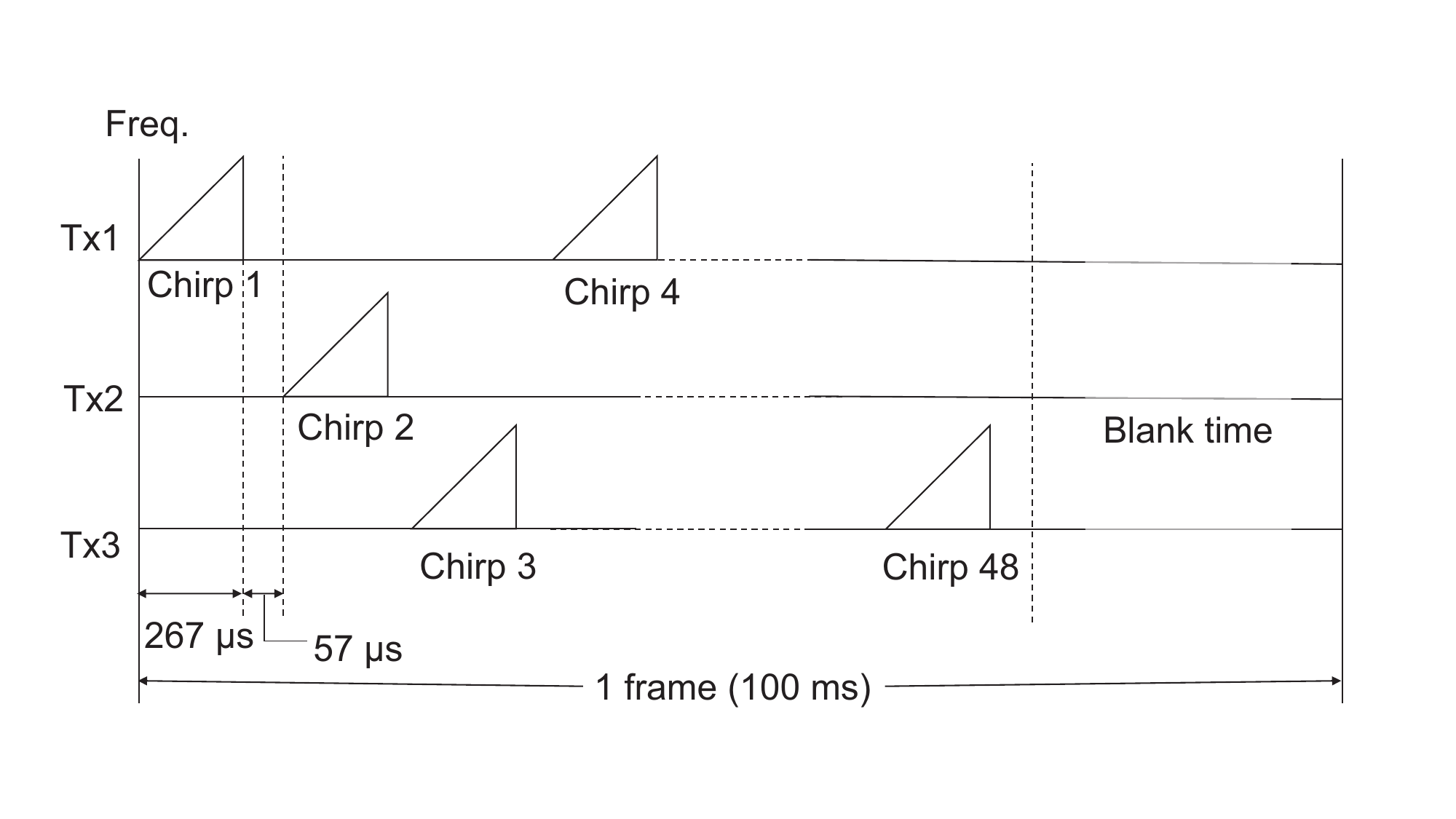}
    \caption{Chirp signal transmission diagram of the radar system in this study.}
    \label{fig:chirptiming}
\end{figure}
Figure \ref{fig:chirptiming} is a diagram of the chirp signal transmission of the radar system used in time-division multiplexing that is applied only to the transmitters and not to the receivers.
The chirp duration was 267 $\mu$s, the interval between consecutive chirps was 57 $\mu$s, and the three transmitting elements thus transmitted signals sequentially with intervals of 324 $\mu$s. After transmitting all 48 chirps, there was a blank time. Note that the time difference between the first and third transmitting signals was 648 $\mu$s, which was sufficiently small for a typical Doppler shift generated by respiration at 79 GHz. We therefore regard the radar signals for the three transmitting elements to be coherent hereafter. 
Note also that the range resolution (43 mm) of the radar system was not identical to the minimum distance between two people in the line-of-sight direction that could be distinguished by the radar. We confirmed through measurements that the minimum separation distance was around 180 mm, which is more than four times the range resolution.

\subsection{Radar Imaging of Multiple People}
\label{subsec:2_B}
\begin{figure}[tb]
    \centering
    \includegraphics[width=0.45\textwidth]{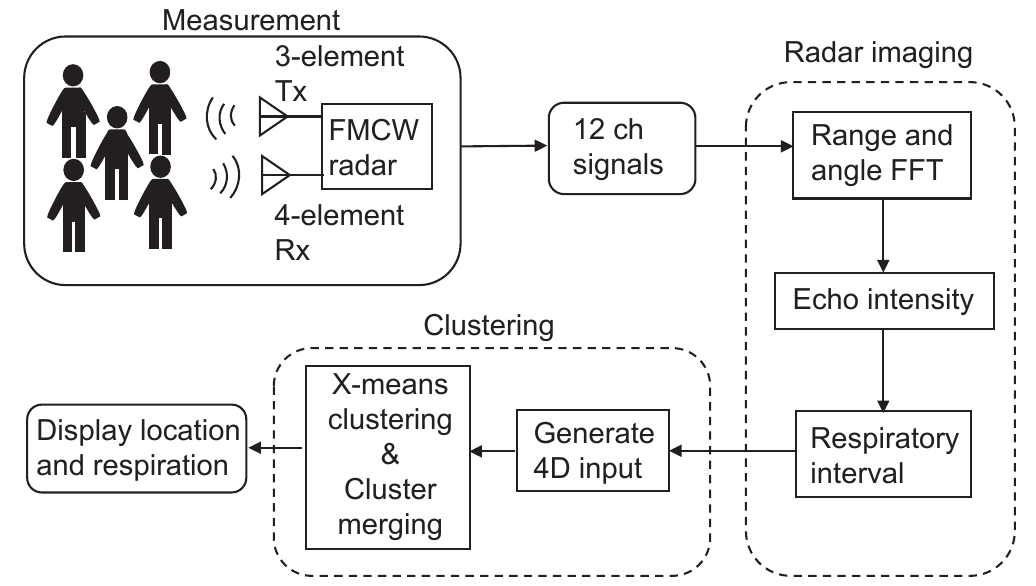}
    \caption{Overview of the radar-based respiratory measurement system for multiple human targets.}
    \label{fig:radar system}
  \end{figure}
Figure \ref{fig:radar system} presents an overview of the entire system. 
The signal received by the FMCW radar can be converted into range-domain data by applying a Fourier transform in the fast-time direction, and it is expressed as $s_k(t,r)$ $(k=0, 1, \cdots, K-1)$. Here, $K=12$ is the number of elements in the virtual array, and variables $t$ and $r$ represent the slow time and range, respectively. The $x$ coordinate of the $k$-th virtual array element is denoted $x_k=k \lambda/2$. Each received signal is multiplied by the calibration correction factor $c_k$ and the Taylor window coefficient $\tau_k$ to obtain the signal vector $\vct{s}(t,r)$, which is defined by
\begin{align}
  \begin{split}
    \vct{s}&(t,r)\\=&[\tau_0c_0s_0(t,r),\tau_1c_1s_1(t,r),\cdots,\tau_{K-1}c_{K-1}s_{K-1}(t,r)]^\mathrm{T},
  \end{split}
\end{align}
where the superscript $\mathrm{T}$ denotes the transpose of the matrix. The beamforming technique is applied to this signal vector to obtain a complex radar image $I_0(t,r,\theta)=\vct{w}^\mathrm{H}(\theta)\vct{s}(t,r)$. Here, the superscript $\mathrm{H}$ represents the complex conjugate transposition of the matrix, and the beamforming weight coefficient vector $\vct{w}(\theta)=[w_0,w_1,\cdots,w_{K-1}]^\mathrm{T}$, with $w_k(\theta)=\mathrm{e}^{-\jj (2\pi x_k/\lambda)\sin\theta}=\mathrm{e}^{-\jj\pi k\sin\theta}$ $(k=0,1,\cdots,K-1)$.

When calculating $I_0(t,r,\theta)$, we discretize $\theta$ into unequal intervals $\theta_0,\theta_1,\cdots,\theta_{N-1}$. By selecting $\theta_n$ such that $\sin\theta_n=n/N$ $(n=0,1,\cdots,N-1)$, we obtain $w_k(\theta_n)=\mathrm{e}^{-\jj\pi kn/N}$. By selecting $N=K$, the equation becomes equivalent to a discrete Fourier transform in the element number direction, allowing the fast Fourier transform algorithm to be used for implementation.

The complex radar image $I_0(t,r,\theta)$ contains static clutter, which is the unwanted response from the stationary objects, in addition to the echoes from the human targets. To remove the static clutter, we subtract the DC component (i.e., the time average) to obtain the complex radar image $I_\mathrm{c}(t,r,\theta)$ in which the static clutter is suppressed:
\begin{align}
  I_\mathrm{c}(t,r,\theta)=I_0(t,r,\theta)-\frac{1}{T_\mathrm{c}} \int_{t-T_\mathrm{c}}^{t}  I_0(\tau,r,\theta)\,\mathrm{d}\tau. 
  \label{eq:DC_removal}
\end{align}
Furthermore, the power of the complex radar image is time-averaged to obtain the radar image $I_\mathrm{P}(t,r,\theta)$: 
\begin{align}
  I_\mathrm{P}(t,r,\theta)=\frac{1}{T_\mathrm{P}} \int_{t-T_\mathrm{P}}^{t}\left|I_\mathrm{c}(\tau,r,\theta)\right|^2\dd\tau,
\end{align}
where $T_\mathrm{c}$ and $T_\mathrm{P}$ are set to 30 and 20 s, respectively.

In most previous studies in this field, $I_\mathrm{P}(t,r,\theta)$ has been used to detect multiple human targets and estimate their locations. The typical approach involves the application of a clustering algorithm in the real space represented by $(r, \theta)$. In this paper, we propose a novel approach in which estimated respiration rate values are exploited to discriminate echoes from human bodies, where the respiration rate is estimated even before estimating the number and locations of the human targets, as described in the next section. This step is the core idea of our proposed system for accurately discriminating the radar echoes from human targets in proximity.

Note that the DC suppression process in Eq. (\ref{eq:DC_removal}) can cause waveform distortion, which may lower the accuracy of the proposed method. The effect of the DC offset on the physiological measurement has been intensively studied \cite{DC1,DC2} and will be considered in our future work.

\subsection{Respiratory Imaging of Multiple People}
In most previous studies, the location $(r_0, \theta_0)$ of each target person is first determined, after which vital signs such as the respiration are extracted from the signal $I_\mathrm{c}(t,r_0,\theta_0)$ corresponding to the location. In our proposed approach, we attempt to estimate the respiratory interval at all the locations $(r,\theta)$ without specifying the location of the human target. As a result, the respiratory interval can be expressed as a function of $t$, $r$, and $\theta$. In this paper, this image is referred to as the respiratory image.

We first estimate the displacement waveform of the target from the phase variation of the reflected waves as
\begin{align}
  d_0(t,r,\theta)=\frac{\lambda}{4\pi}\angle I_\mathrm{c}(t,r,\theta).
  \label{eq:disp0}
\end{align}
We next apply a bandpass filter to $d_0(t,r,\theta)$ and obtain $d(t,r,\theta)$, where the cut-off frequencies of the filter correspond to 11 and 51 s.
This bandpass filter suppresses the low-frequency component, which mainly contains body movement, and the high-frequency component, which mainly contains noise.

Finally, the respiratory interval at time $t$ is determined as a function of $r$ and $\theta$ as $\tau_\mathrm{r}(t,r,\theta)$ by solving an optimization problem:
\begin{align}
  \tau_{\mathrm{r}}(t,r,\theta)=\arg\min_{0<\tau\leq T_0} f_{t,r,\theta}(\tau),
  \label{eq:resp_period}
\end{align}
where we set $T_0=8.0$ s, and the cost function $f_{t,r,\theta}(\tau)$ is expressed as
\begin{align}
  \begin{split}
    f_{t,r,\theta}(\tau)&=  \frac{1}{2T_0} \int_{t-2T_0}^{t} |d(t',r,\theta)-d(t'+\tau,r,\theta)\vert^2\\
    &+|d(t',r,\theta)-d(t'-\tau,r,\theta)\vert^2 \,\mathrm{d}t'.
  \end{split}
  \label{eq:resp_estimate}
\end{align}
The resultant image $\tau_{\mathrm{r}}(t,r,\theta)$ is called respiratory image that is used in our proposed clustering method. In the implementation of the above algorithm, $\tau_\mathrm{r}(t,r,\theta)$ does not represent a respiratory interval when there is no echo from a human target at $(r, \theta)$. The computational load can be reduced by setting a certain threshold and calculating the respiratory image $\tau_\mathrm{r}(t,r,\theta)$ only for $(t, r, \theta)$ with the radar image exceeding the threshold.

\section{Estimation of the Number and Locations of Multiple Human Targets with Clustering}
\subsection{Conventional x-means Clustering with an Unknown Number of Clusters}
Numerous studies on radar imaging and tracking used clustering algorithms to separate echoes from multiple targets, including human bodies \cite{20,21,22,24,26,25,27,28,29,30,31,32,33,34,35}. 
In this study, we also adopted the x-means clustering algorithm \cite{Pelleg}, which is an extension of the well-known $k$-means algorithm, because the x-means algorithm does not require the number of clusters whereas the $k$-means algorithm does. The x-means algorithm is therefore suitable for situations where the number of targets is unknown, as in this study. The x-means algorithm includes the following steps \cite{Pelleg}.
\begin{enumerate}
  \item  Let cluster $C$ represent all the input data (the point cloud).
  \item  Process the point cloud $\vct{x}\in C$ using the $k$-means method with $k=2$ to generate clusters $C_1$ and $C_2$.
  \item Calculate the Bayesian information criterion for cluster $C$ and for the two segmented models $C_1$ and $C_2$. Let the former criterion be $b$ and the latter be $b'$.
  \item  If $b\leq b'$, adopt cluster $C$ before the split.
  \item If $b>b'$, adopt the split clusters $C_1$ and $C_2$ and return to step (2), setting $C_1\rightarrow C$. Then repeat step (2) setting $C_2\rightarrow C$.
\end{enumerate}
In the conventional approach, the initial point cloud $C$ is generated from the radar image $I_\mathrm{P}(t,r,\theta)$ alone, whereas in the next section, we introduce a respiratory-space representation as the point cloud $C$ that contains information on the respiratory characteristics estimated from the complex radar image $I_\mathrm{c}(t,r_0,\theta_0)$. At time $t$, a point cloud is generated in the $r$-$\theta$ domain from the radar image $I_\mathrm{P}(t,r,\theta)$ and clustered using the x-means algorithm, where the number of targets is obtained from the estimated number of clusters. The target locations are estimated from the centroid of each cluster. We refer to the clustering method explained in this section as conventional two-dimensional (2D) clustering.

\subsection{Proposed Respiratory-space Clustering}
In this section, we propose a new clustering method, respiratory-space clustering, to separate the radar echoes of multiple breathing human targets and determine the target number and locations. The proposed method estimates the respiratory interval for each location $(r, \theta)$ prior to clustering, increasing the dimensionality of the input point cloud by converting the respiratory intervals into virtual spatial coordinates. The resulting high-dimensional space is called the respiratory space. The point cloud in the respiratory space is then processed using the x-means algorithm to estimate the number and locations of people.

The range $r$ and azimuth $\theta$ are discretized and denoted $r_l$ and $\theta_n$ $(l=1,2,\cdots,L; n=1,2,\cdots,N)$, respectively. The sampling interval of the range $r$ is a constant value $\Delta r$, while the angles $\theta$ are sampled at unequal intervals, as described in Section \ref{subsec:2_B}. In this paper, we define the four-dimensional vector ${\vct{x}_{l,n}}$ in the respiratory space at time $t$ as
\begin{align}
    {\vct{x}_{l,n}}=[r_l, \theta_n, \tau_\mathrm{r}(t,r_l,\theta_n), \tau_\mathrm{r}(t-T_\mathrm{cy},r_l,\theta_n)]^\mathrm{T},
    \label{eq:4Dvector}
\end{align}
where $T_\mathrm{cy}=6.0$ s. This four-dimensional vector contains the respiratory intervals calculated at time $t$ and $t-T_\mathrm{cy}$ in addition to the physical coordinates $(r,\theta)$. To suppress the effects of noise and random components, we use respiratory intervals averaged for time $T_\mathrm{cy}$, which is approximately a typical respiratory interval, and apply a median filter sized $3\times 4$.

When generating the point cloud to be input to the clustering algorithm, it is preferable that the point density is high for $(r, \theta)$ having a large value in the radar image $I_\mathrm{P}$. The number of point cloud points $\rho _{l,n}\in \mathbb{Z}$ used for clustering is given by 
\begin{align}
  \rho_{l,n}=[\alpha r_l I_\mathrm{P}(t,r_l,\theta_n)],
\end{align}
where $\alpha$ is a constant, $[\;\cdot\;]$ represents the integer closest to the value of the real number, and $r_l$ is a multiplicative factor that compensates for propagation loss. Using this formula, points were generated in four-dimensional space. This process is repeated for all $(l,n)$ to generate the point cloud, which is then passed as input to the x-means algorithm for clustering.

\subsection{Proposed Cluster Merging Method}
In the proposed method, two or more clusters may be generated erroneously for the same target. In such cases, the corresponding clusters would be located close to each other. Such an erroneous clustering can be detected by considering the typical size of the human body $D$, where we set $D=0.6$ m empirically. In this paper, we propose the following steps for merging multiple clusters that have been incorrectly split. Let us assume that $M$ clusters $C^1,C^2\cdots, C^M$ are generated using the method proposed in the previous section.  
\begin{enumerate}
    \item Calculate the centroid of each cluster. The centroids, denoted by ${\bm c_1},{\bm c_2},\cdots,{\bm c_M}$, are points on the $x$--$y$ plane, which are the Cartesian coordinates converted from the $r$--$\theta$ plane.
    \item Calculate the distance $g_{i,j}=\| {\bm c_i}-{\bm c_j}\Vert \;\;(i,j=1,2,\cdots,M)$ between each centroid pair.
    \item If $D>\min_{i,j} g_{i,j}$, merge clusters $C^{i^*}$ and $C^{j^*}$ letting $(i^*,j^*)=\arg\min_{i,j} g_{i,j}$ and return to step (1) unless $D\leq\min_{i,j} g_{i,j}$.
\end{enumerate}
The final number of clusters generated by the above process is output as an estimate of the number of participants. After applying this cluster merging method, the representative position of each cluster is determined as the position where the value of the radar image $I_\mathrm{P}$ is a maximum within the cluster area.

\section{Performance Evaluation of the Proposed Respiratory-space Clustering}
  \subsection{Overview of the Experimental Setup}
  \begin{figure}[t]
    \centering
    \includegraphics[width=0.45\textwidth]{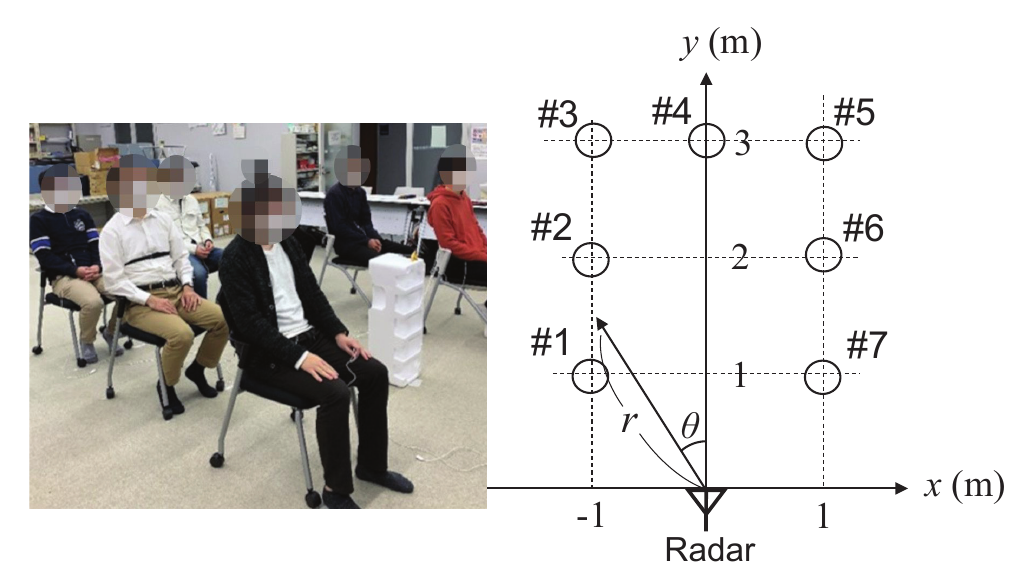}
    \\(a)\hspace{30mm}(b)\vspace{3mm}
    \caption{Experiment 1: Experimental validation with seven participants (a) and their seating positions (b).}
    \label{fig:sokutei1}
  \end{figure}
  \begin{figure}[t]
    \centering
    \includegraphics[width=0.45\textwidth]{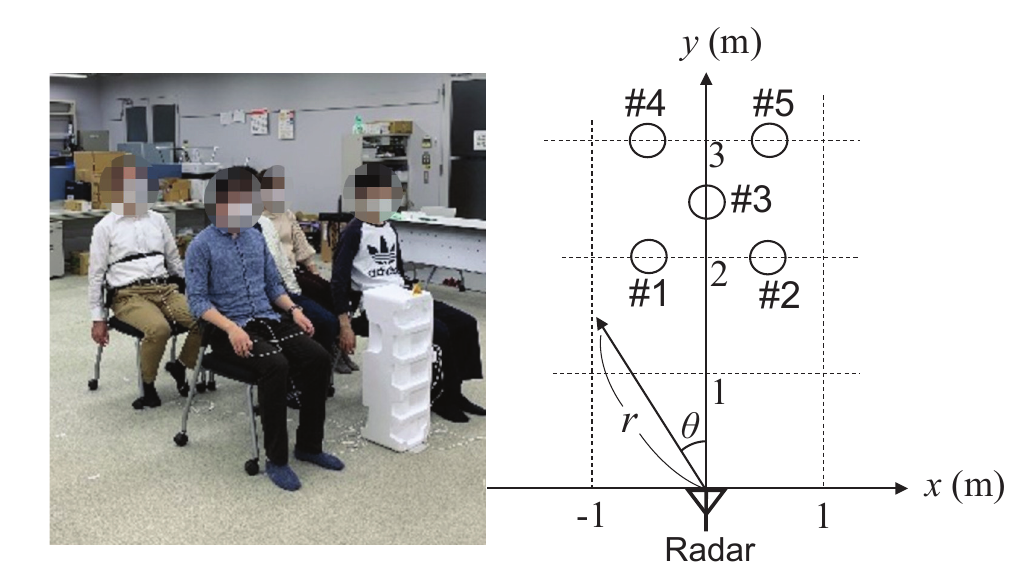}
    \\(a)\hspace{30mm}(b)\vspace{3mm}
    \caption{Experiment 2: Experimental validation with five participants in proximity (a) and their seating positions (b).}
    \label{fig:sokutei2}
  \end{figure}
  We conducted experiments to evaluate the performance of the proposed respiratory-space clustering method for the respiratory measurement of multiple human targets. The participants were seated and stationary. Experiment 1 assumed seven seated participants spaced at approximately 1-m intervals (Fig. \ref{fig:sokutei1}), whereas experiment 2 assumed five seated participants spaced at approximately 0.7-m intervals (Fig. \ref{fig:sokutei2}). The seven participants in experiment 1 were seated in a U-shaped arrangement, as shown in the right panel of Fig. \ref{fig:sokutei1}. The four participants in experiment 2 were seated at the corners of a 1-m square, with the fifth participant at the center, as depicted in the right panel of Fig. \ref{fig:sokutei2}. In both experiments, the coordinates of the center of the radar array were used as the origin. The measurement time was 120 s in both experiments, and the participants were instructed to remain still and breathe naturally during the measurement. We simultaneously measured the respiratory rate using belt-type sensors worn on the chests of the participants to quantitatively evaluate the accuracy of the respiratory rate estimated using the proposed radar-based approach.

  \subsection{Evaluation of the Number and Location Estimates}
  \subsubsection{Experiment 1}
\begin{figure}[t]
    \centering
    \includegraphics[width=0.35\textwidth]{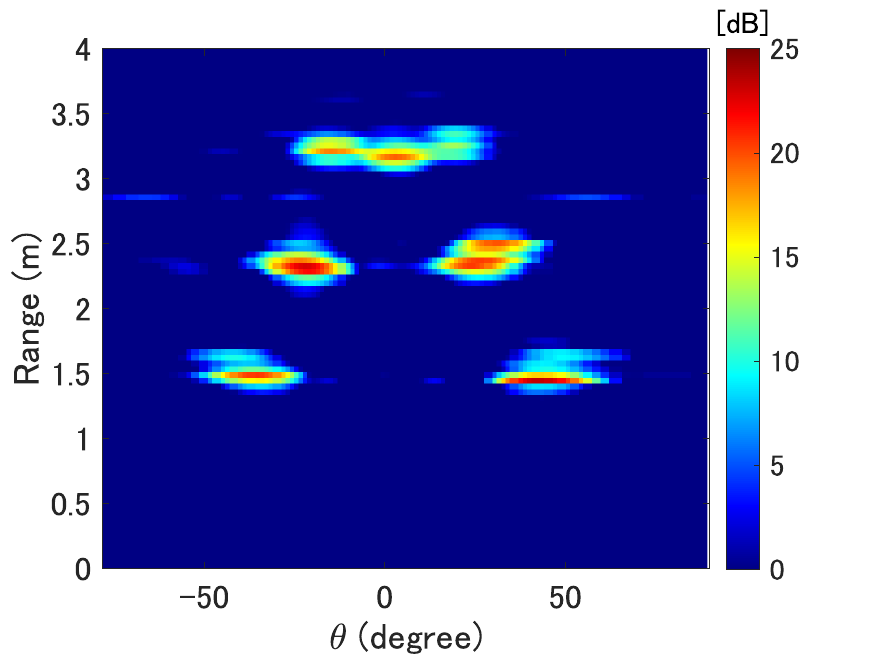}
    \caption{Example of the radar image $I_\mathrm{P}(t,r,\theta)$ (experiment 1).}
    \label{fig:echointensity1}
  \end{figure}

\begin{figure}[t]
    \centering
    \includegraphics[width=0.35\textwidth]{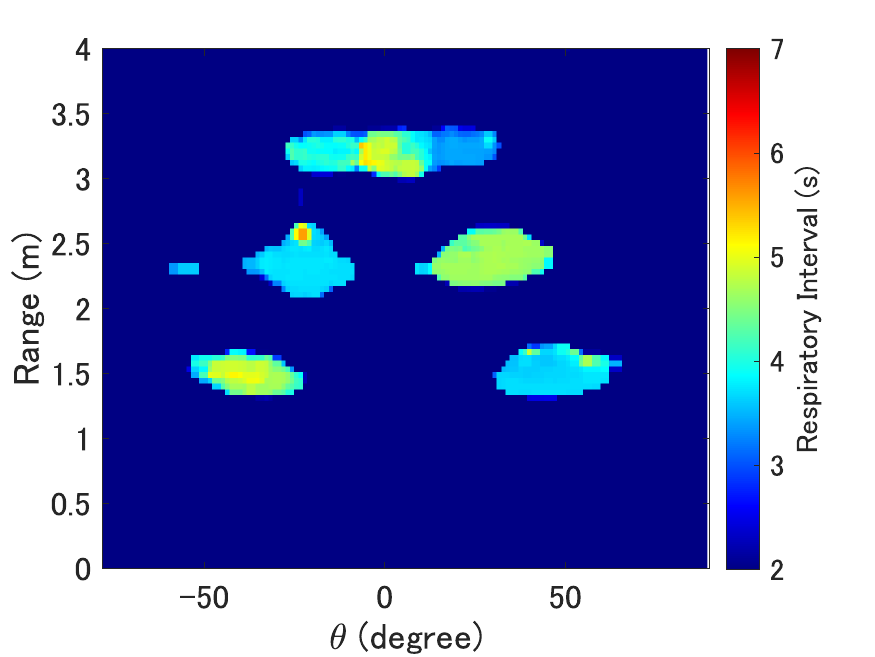}
    \caption{Example of the respiratory image $\tau_{\mathrm{r}}(t,r,\theta)$ (experiment 1).}
    \label{fig:respiratory_interval1}
  \end{figure}

  Figure \ref{fig:echointensity1} presents an example of the radar image $I_\mathrm{P}(t,r,\theta)$ captured in experiment 1 with the seven participants. Note that $I_\mathrm{P}(t,r,\theta)$ has been normalized with the noise floor. In this figure, strong responses are seen at the approximate seating positions of the seven participants. Figure \ref{fig:respiratory_interval1} shows an example of the respiratory image $\tau_{\mathrm{r}}(t,r,\theta)$, where different respiratory intervals are observed at the seating positions.
  
  
\begin{figure}[t]
    \centering
    \includegraphics[width=0.35\textwidth]{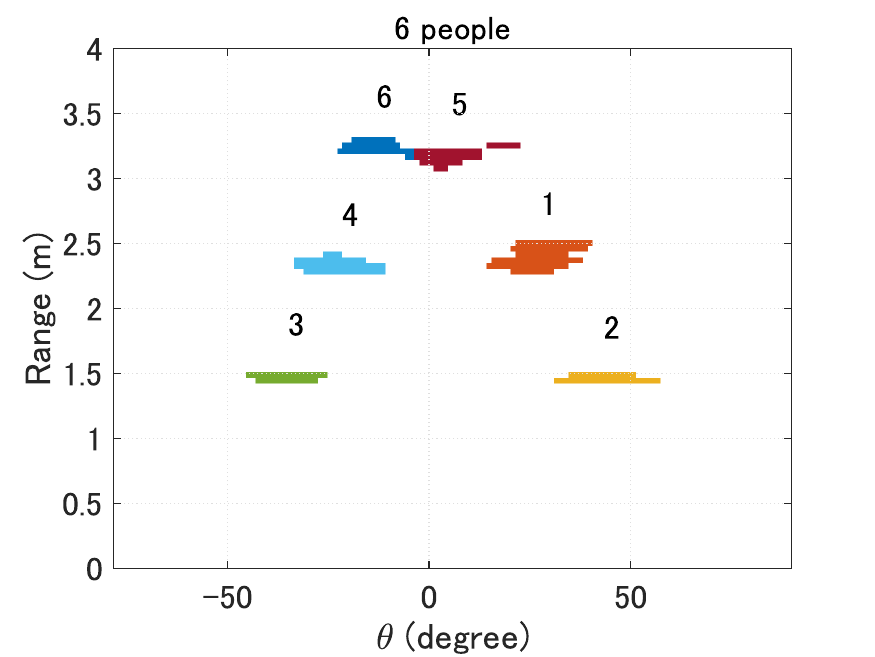}
    \caption{Estimation using conventional 2D clustering (experiment 1).}
    \label{fig:2Dcluster1}
  \end{figure}

  We next present the clusters obtained using the conventional and proposed clustering methods. We applied conventional clustering in 2D space $(r,\theta)$ and the proposed clustering in four-dimensional (4D) space $(r,\theta,\tau_1,\tau_2)$. Figure \ref{fig:2Dcluster1} shows the clusters obtained using the conventional 2D clustering. Note that the cluster numbers do not necessarily match the participant numbers shown in Fig. \ref{fig:sokutei1}. Although four participants within a range of 2.5 m were correctly clustered, three participants at ranges greater than 3 m were incorrectly clustered into two clusters; i.e., the number of people was estimated incorrectly. When the targets are located far from the radar antennas, multiple targets in the $r$--$\theta$ space become closer in the angle direction, making them more difficult to separate.
  
\begin{figure}[t]
    \centering
    \includegraphics[width=0.35\textwidth]{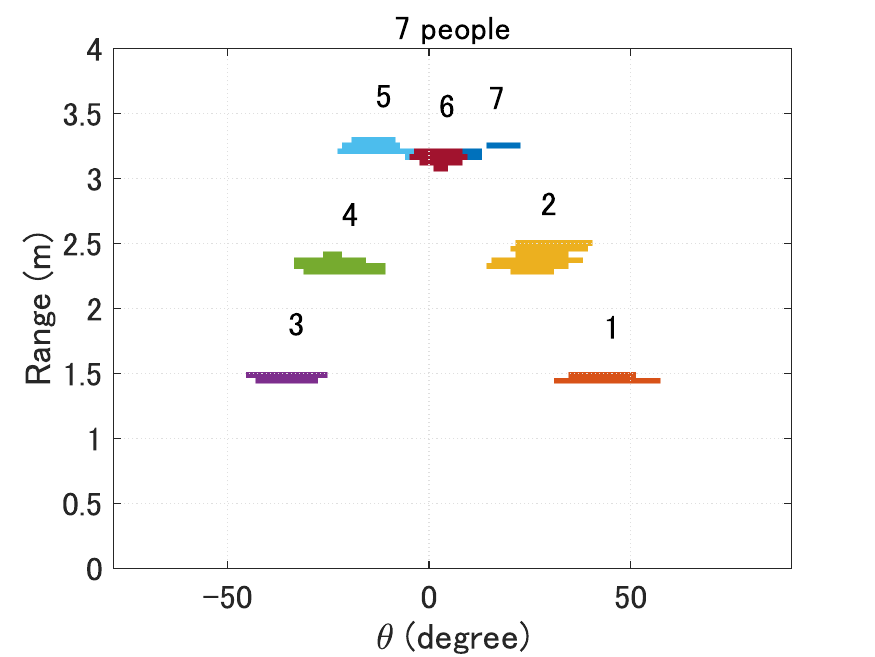}
    \caption{Estimation using the proposed respiratory space clustering (experiment 1).}
    \label{fig:4Dcluster1}
  \end{figure}

  Figure \ref{fig:4Dcluster1} shows the clusters generated using the proposed 4D respiratory-space clustering for the same time instance as in Fig. \ref{fig:2Dcluster1}; all the participants are correctly clustered regardless of the range. Because the x-means algorithm is initialized with random numbers, the estimated number of targets depends on the random numbers. To evaluate the accuracy in estimating the number of targets, we ran the algorithm with 100 random seeds. Because the proposed method performed clustering 11 times over a measurement duration of 120 s, the total number of clustering trials was 1100 for 100 seeds. Among the 1100 clustering trials, the number of targets was correctly estimated 100 and 1073 times using the conventional 2D clustering and the proposed 4D clustering, respectively. These numbers correspond to 9.1\% and 97.6\% of accuracies in the estimation of the number of targets. Thus, high-dimensional clustering, which incorporates the respiratory intervals in the input vector, is an effective approach for accurately estimating the number of human targets.
  
\begin{figure}[t]
    \centering
    \includegraphics[width=0.35\textwidth]{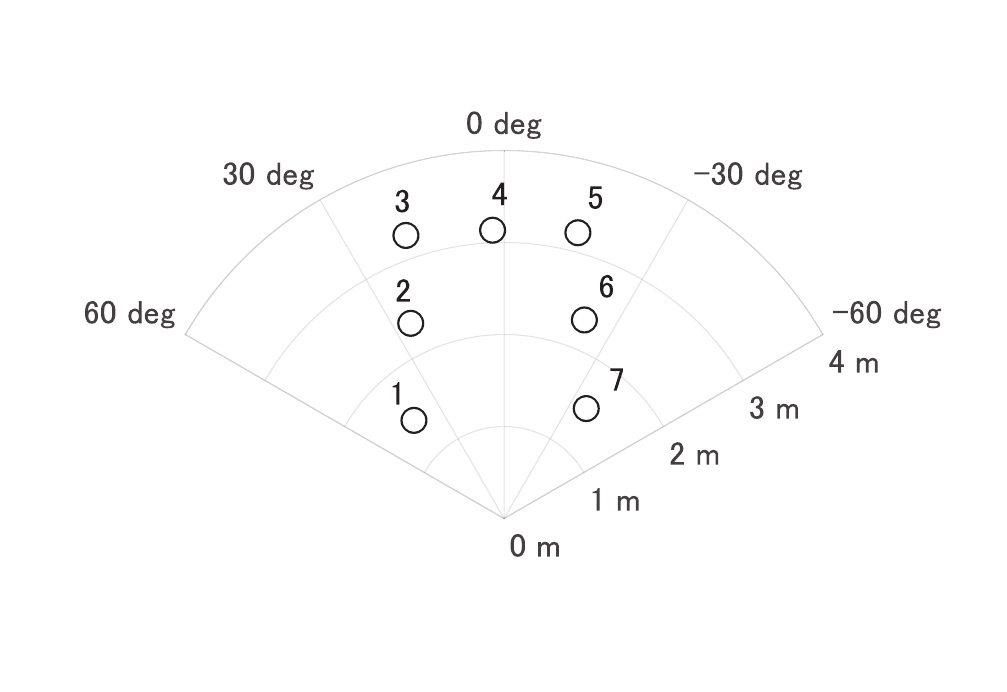}
    \caption{Subject locations estimated using the proposed respiratory-space clustering technique (experiment 1).}
    \label{fig:Position1}
  \end{figure}

  Finally, we present the locations of the participants estimated using the proposed method. The participant locations estimated using the proposed respiratory-space clustering technique are depicted in Fig. \ref{fig:Position1}. Note that Fig. \ref{fig:Position1} shows the locations averaged over the measurement time, where the participant numbers match the numbering shown in Fig. \ref{fig:sokutei1}. The estimated coordinates of the seven participants appear to be correct, and they approximately match the image on the right of Fig. \ref{fig:sokutei1}. However, as the actual participant locations cannot be determined precisely and it is difficult to determine the part of the body contributing to the echo reflection because of the complex shape of the human body, it was not possible to evaluate the accuracy of the estimated locations.
  
  \subsubsection{Experiment 2}
\begin{figure}[t]
    \centering
    \includegraphics[width=0.35\textwidth]{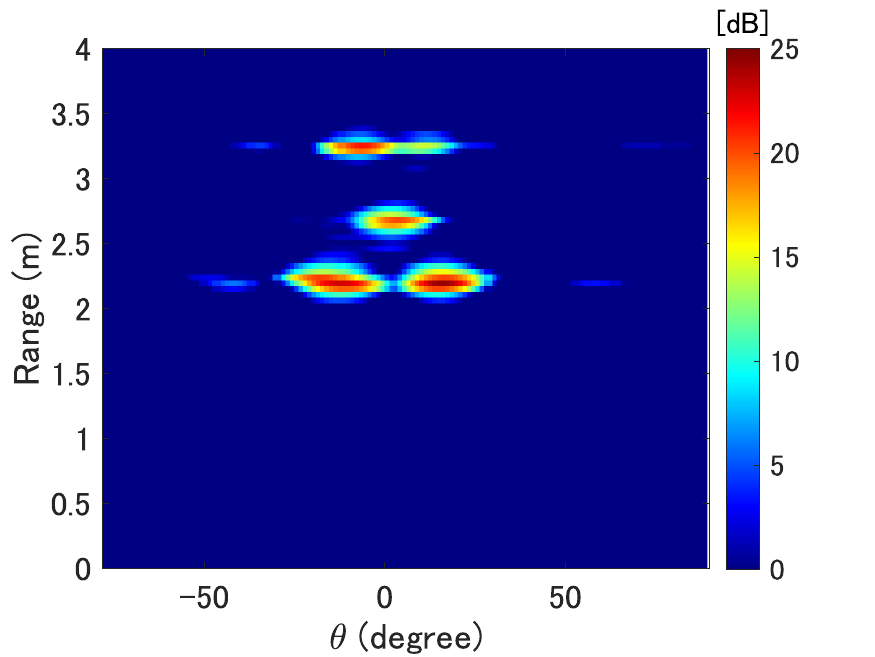}
    \caption{Example of the radar image $I_\mathrm{P}(t,r,\theta)$ (experiment 2).}
    \label{fig:echointensity2}
  \end{figure}

\begin{figure}[t]
    \centering
    \includegraphics[width=0.35\textwidth]{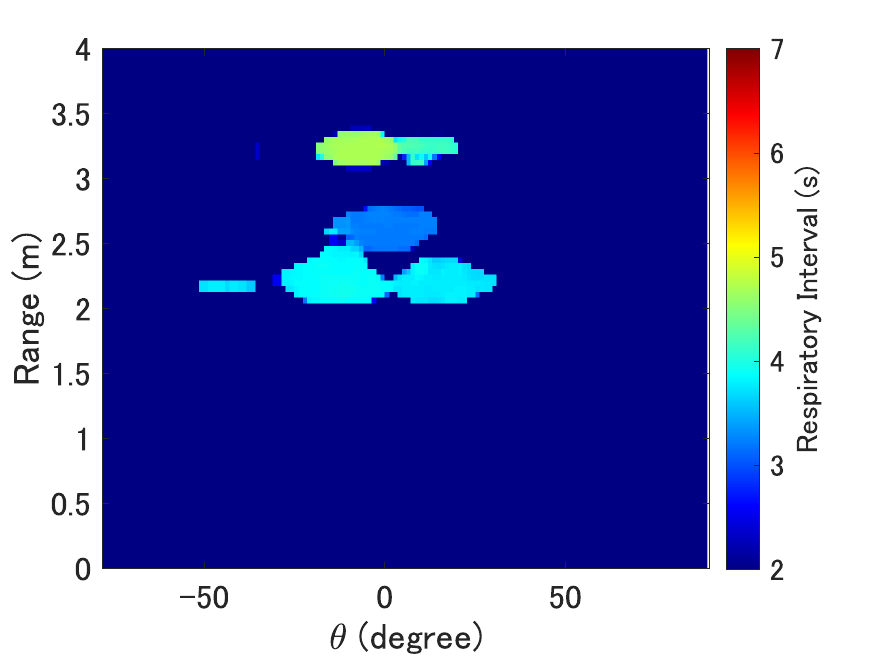}
    \caption{Example of the respiratory image $\tau_{\mathrm{r}}(t,r,\theta)$ (experiment 2).}
    \label{fig:respiratory_interval2}
  \end{figure}

  Figures \ref{fig:echointensity2} and \ref{fig:respiratory_interval2} show examples of the radar image $I_\mathrm{P}(t,r,\theta)$ and respiratory image $\tau_{\mathrm{r}}(t,r,\theta)$ in experiment 2 with five participants. Participants 1 and 2 had similar respiratory intervals by coincidence, whereas the other three participants had different respiratory intervals.
  
\begin{figure}[t]
    \centering
    \includegraphics[width=0.35\textwidth]{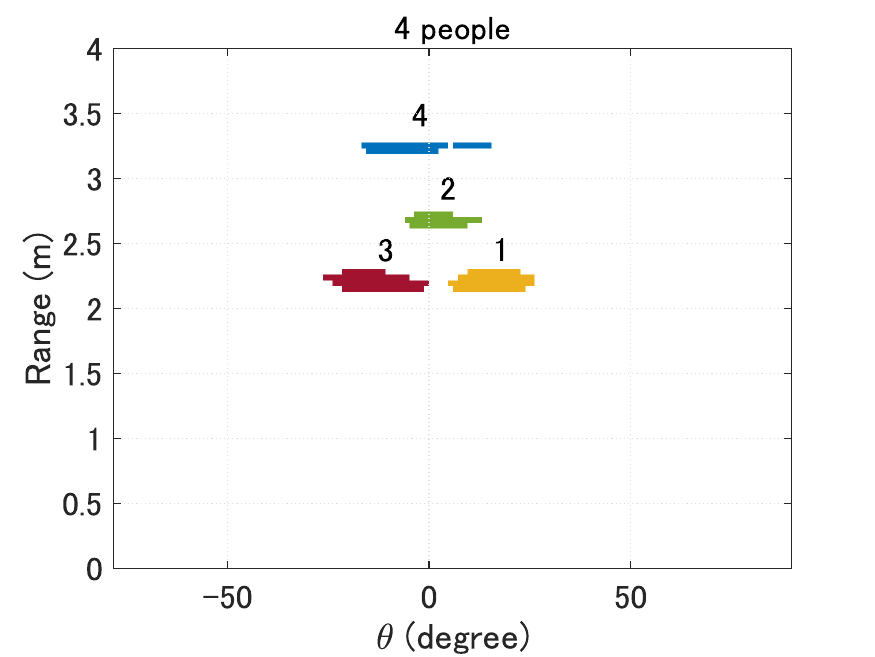}
    \caption{Clusters estimated using conventional 2D clustering (experiment 2).}
    \label{fig:2Dcluster2}
  \end{figure}

  As for experiment 2, we compare the results of conventional 2D clustering with those of the proposed respiratory-space clustering. Figure \ref{fig:2Dcluster2} shows the clusters estimated using conventional 2D clustering. Three participants within a range of 3 m from the radar were separated correctly, but the two participants located at a range of approximately 3.5 m were incorrectly grouped into one cluster. As in experiment 1, distant participants were more difficult to segregate in the angular direction.
  
\begin{figure}[t]
    \centering
    \includegraphics[width=0.35\textwidth]{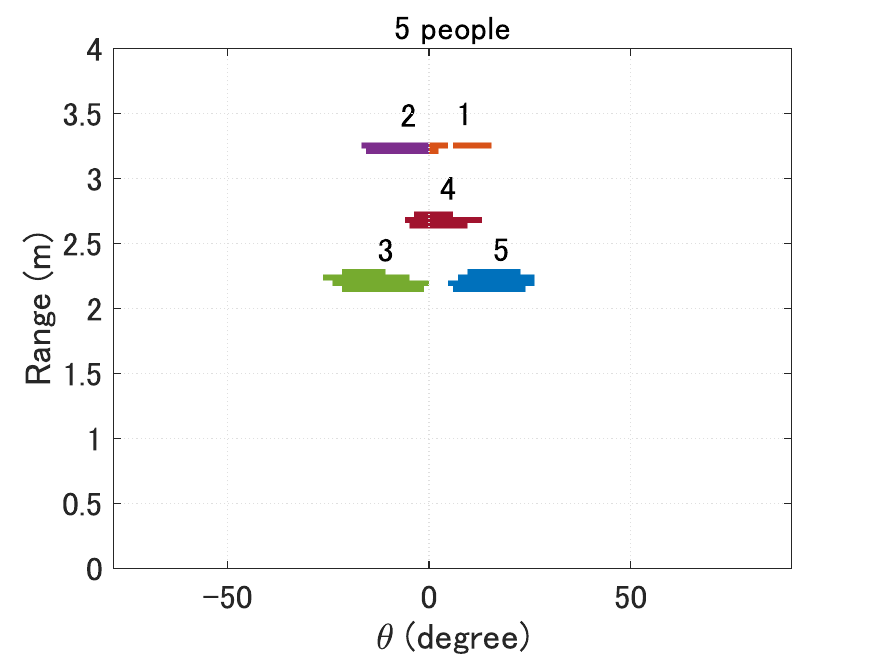}
    \caption{Clusters estimated using the proposed respiratory-space clustering (experiment 2).}
    \label{fig:4Dcluster2}
  \end{figure}

\begin{figure}[t]
    \centering
    \includegraphics[width=0.35\textwidth]{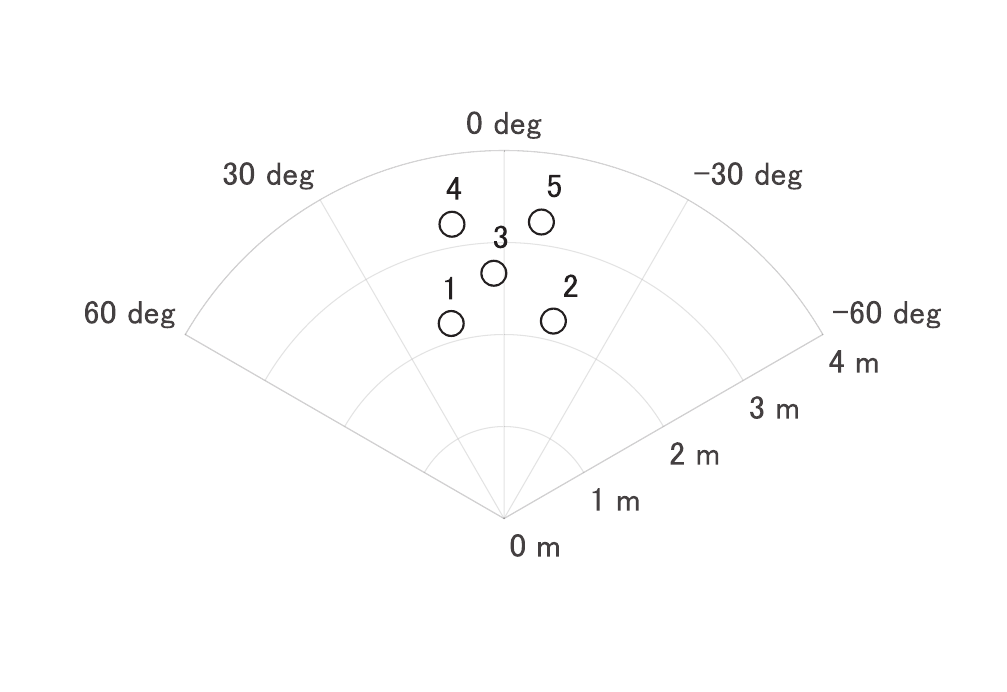}
    \caption{Participant locations estimated using the proposed respiratory-space clustering technique (experiment 2).}
    \label{fig:Position2}
  \end{figure}

  Figure \ref{fig:4Dcluster2} presents the clusters obtained using the proposed respiratory-space clustering at the same time instance as in Fig. \ref{fig:2Dcluster2}. The proposed method correctly separates the two participants at a range of approximately 3.5 m and accurately estimates the number of people. As mentioned above, clustering was performed 11 times in total during the measurement time. Among the 1100 clustering trials for 100 seeds, the number of targets was correctly estimated 200 and 1096 times using the conventional 2D clustering and the proposed 4D clustering, respectively. These numbers correspond to 18.2\% and 99.6\% of accuracies in the estimation of the number of targets. Thus, as in experiment 1, it was demonstrated that high-dimensional clustering is effective in estimating the accurate number of target people.
  Finally, we present the estimated locations of the human targets. The target locations estimated using the respiratory-space clustering technique are shown in Fig. \ref{fig:Position2}; the five participants approximately match the image on the right of Fig. \ref{fig:sokutei2}.

  \subsection{Performance Evaluation of the Proposed Method for Different Layouts of People}
  \begin{figure}[t]
      \centering
      \begin{tabular}{c}
          \begin{minipage}[t]{\hsize}
              \centering
              \includegraphics[width=\textwidth]{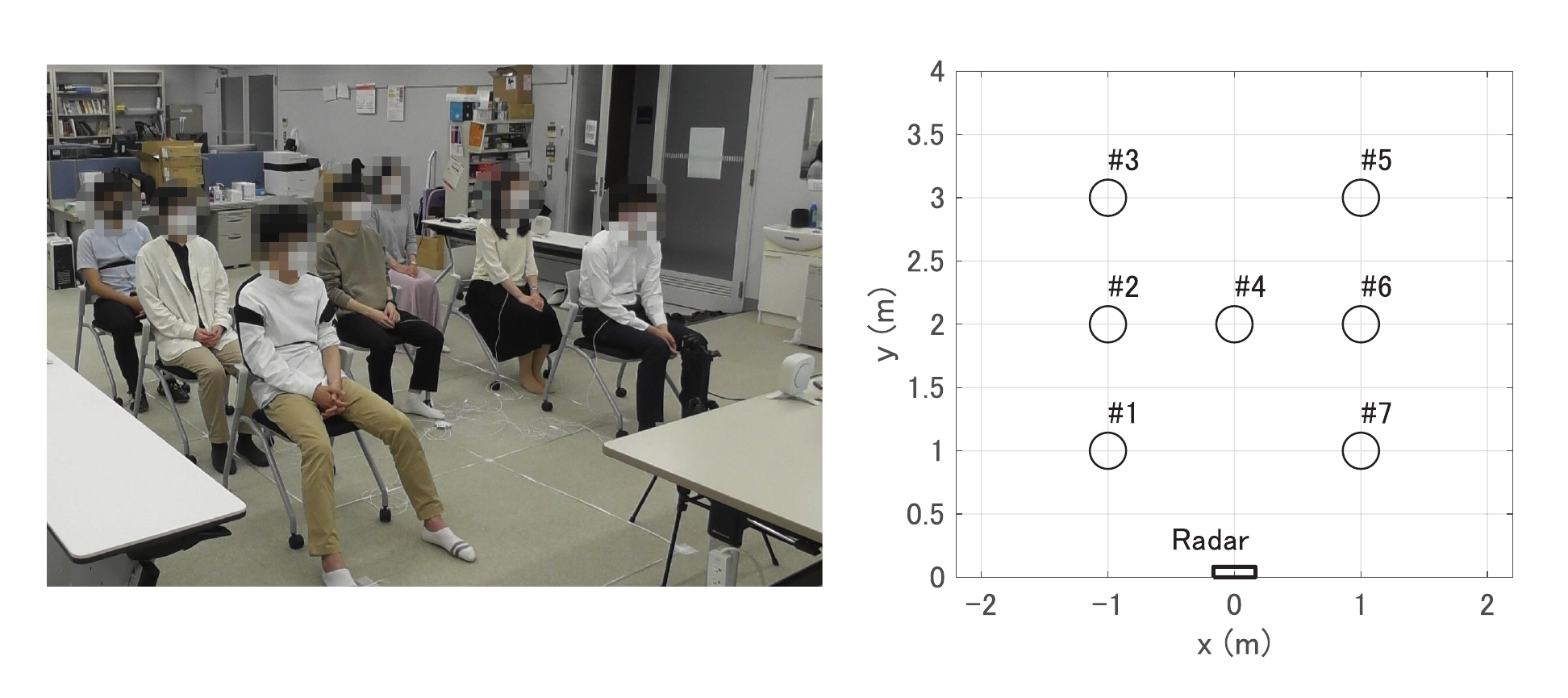}
              \\\hspace{5mm}(a)\hspace{40mm}(b)\vspace{3mm}
          \end{minipage}
      \end{tabular}
      \\
      \begin{tabular}{c}
        \begin{minipage}[t]{\hsize}
            \centering
            \includegraphics[width=\textwidth]{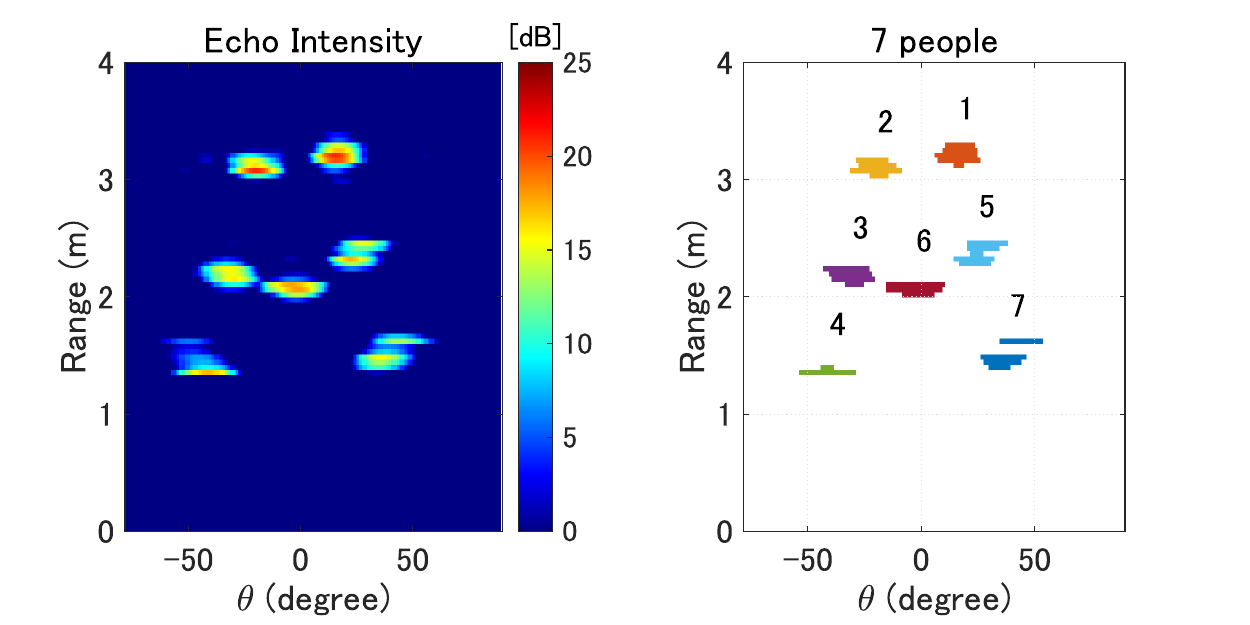}
            \\(c)\hspace{40mm}(d)\vspace{3mm}
        \end{minipage}
    \end{tabular}
    \caption{Experiment 3 with seven participants (a), their seating positions (b), the radar image (c), and clusters estimated using the proposed respiratory-space clustering (d).}
    \label{fig:radar1_case2}
  \end{figure}  
  \begin{figure}[t]
    \centering
    \begin{tabular}{c}
        \begin{minipage}[t]{\hsize}
            \centering
            \includegraphics[width=\textwidth]{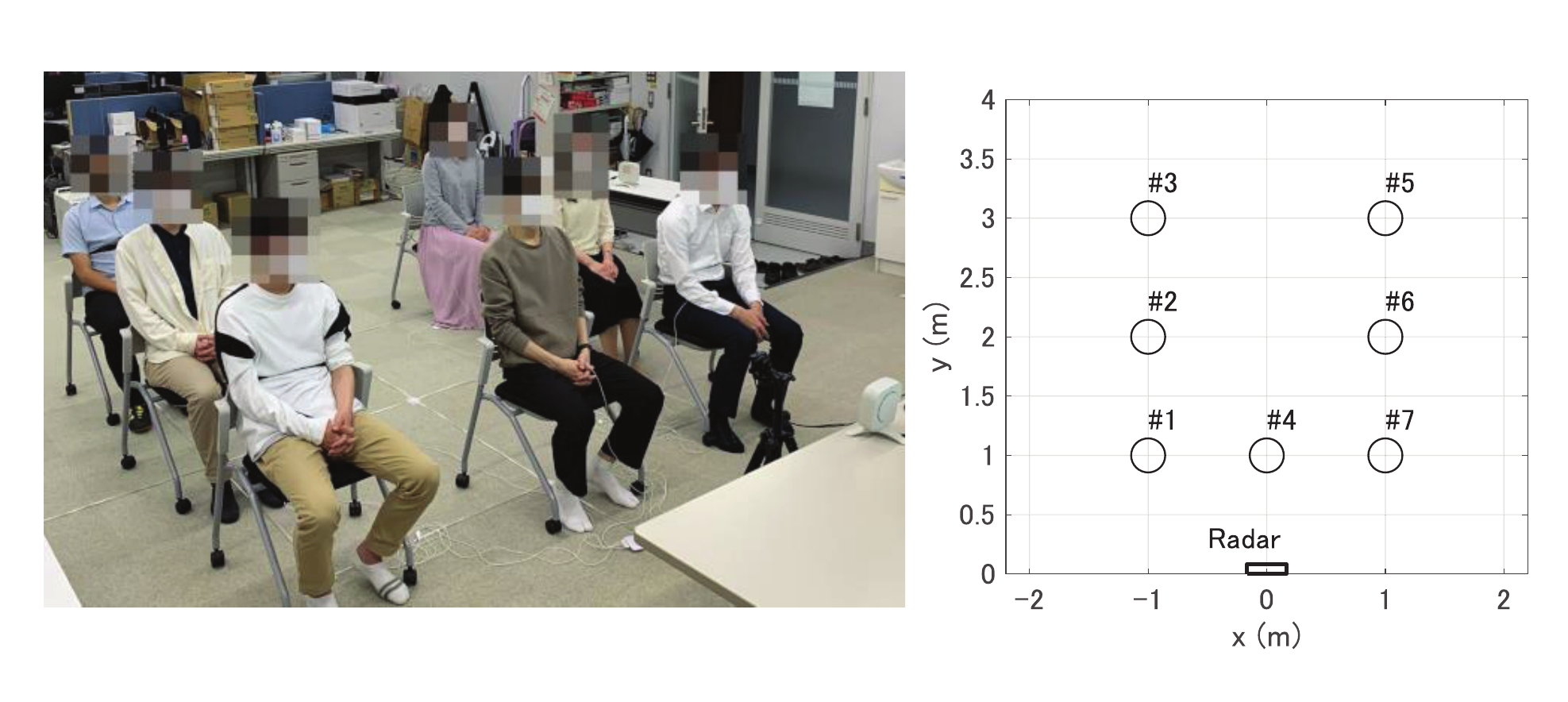}
            \\\hspace{5mm}(a)\hspace{40mm}(b)\vspace{3mm}
        \end{minipage}
    \end{tabular}
    \\
    \begin{tabular}{c}
      \begin{minipage}[t]{\hsize}
          \centering
          \includegraphics[width=\textwidth]{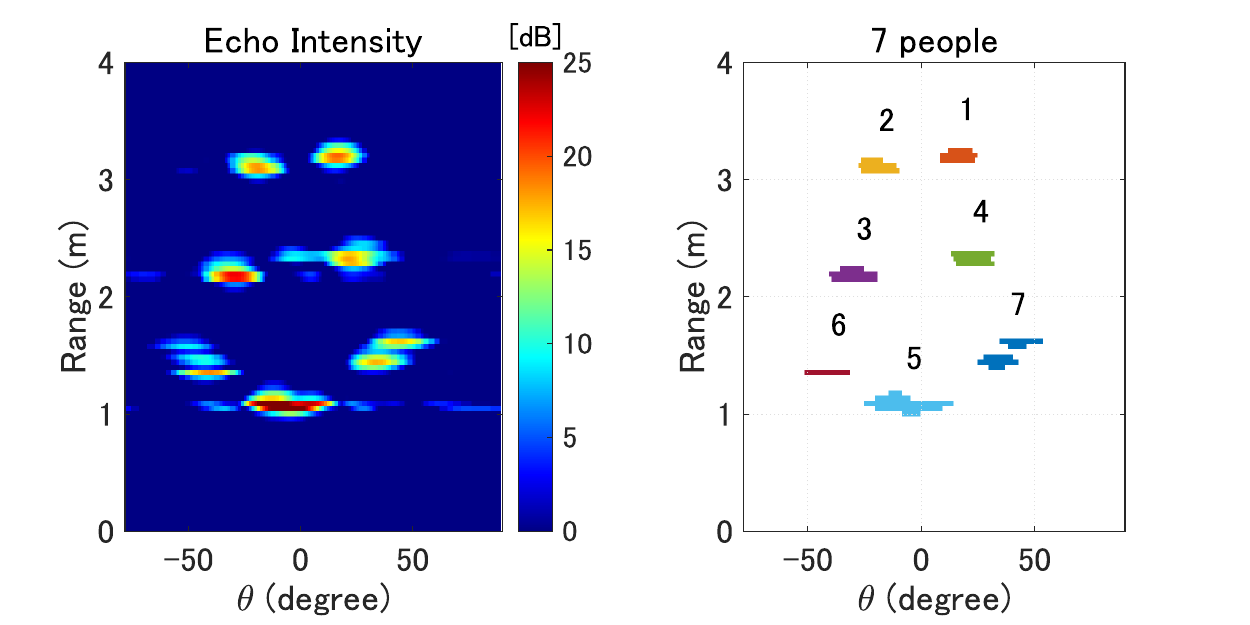}
          \\(c)\hspace{40mm}(d)\vspace{3mm}
      \end{minipage}
  \end{tabular}
  \caption{Experiment 4 with seven participants (a), their seating positions (b), the radar image (c), and clusters estimated using the proposed respiratory-space clustering (d).}
  \label{fig:radar1_case3}
\end{figure}  
\begin{figure}[t]
    \centering
    \begin{tabular}{c}
        \begin{minipage}[t]{\hsize}
            \centering
            \includegraphics[width=\textwidth]{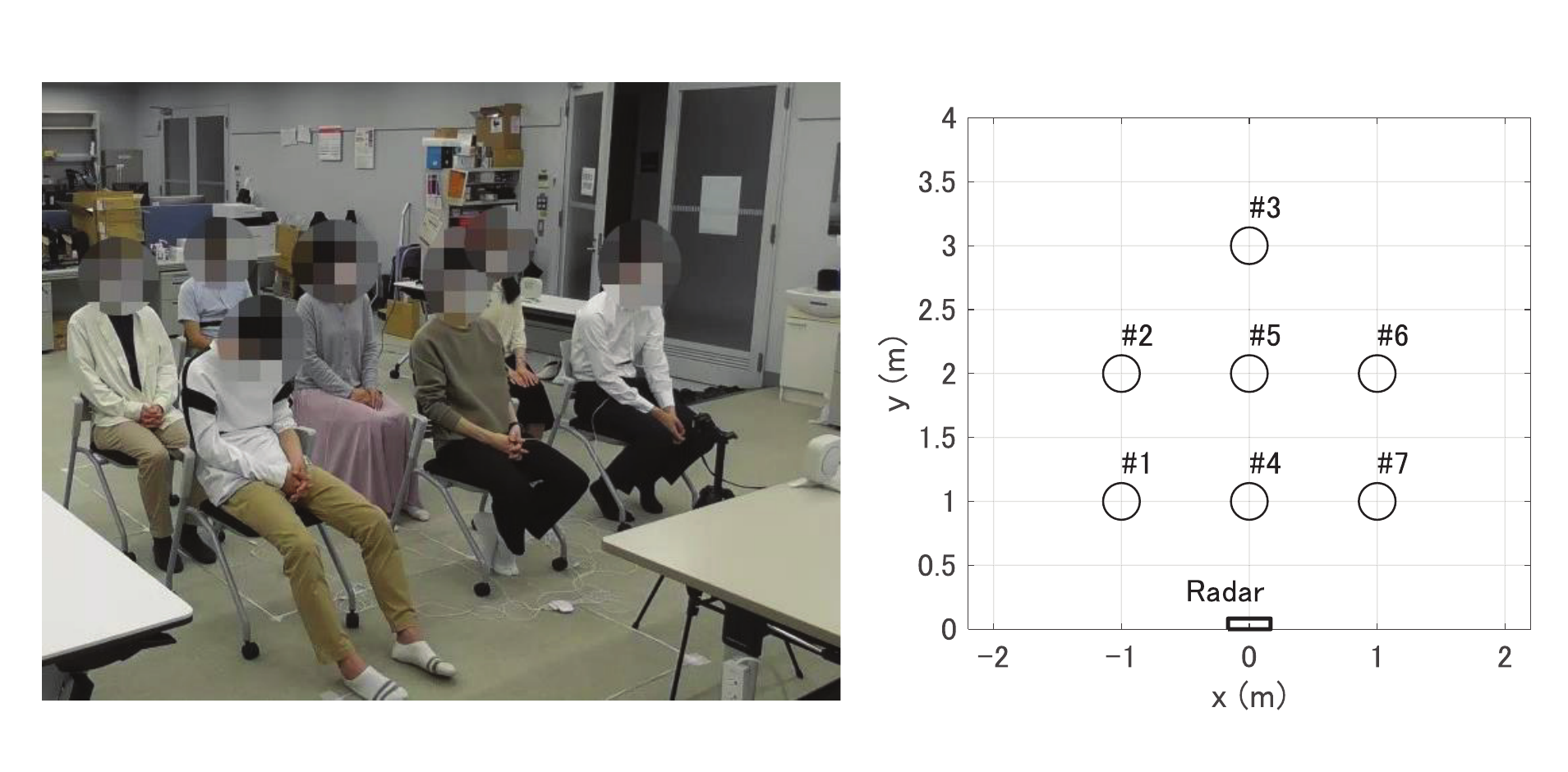}
            \\\hspace{5mm}(a)\hspace{40mm}(b)\vspace{3mm}
        \end{minipage}
    \end{tabular}
    \\
    \begin{tabular}{c}
      \begin{minipage}[t]{\hsize}
          \centering
          \includegraphics[width=\textwidth]{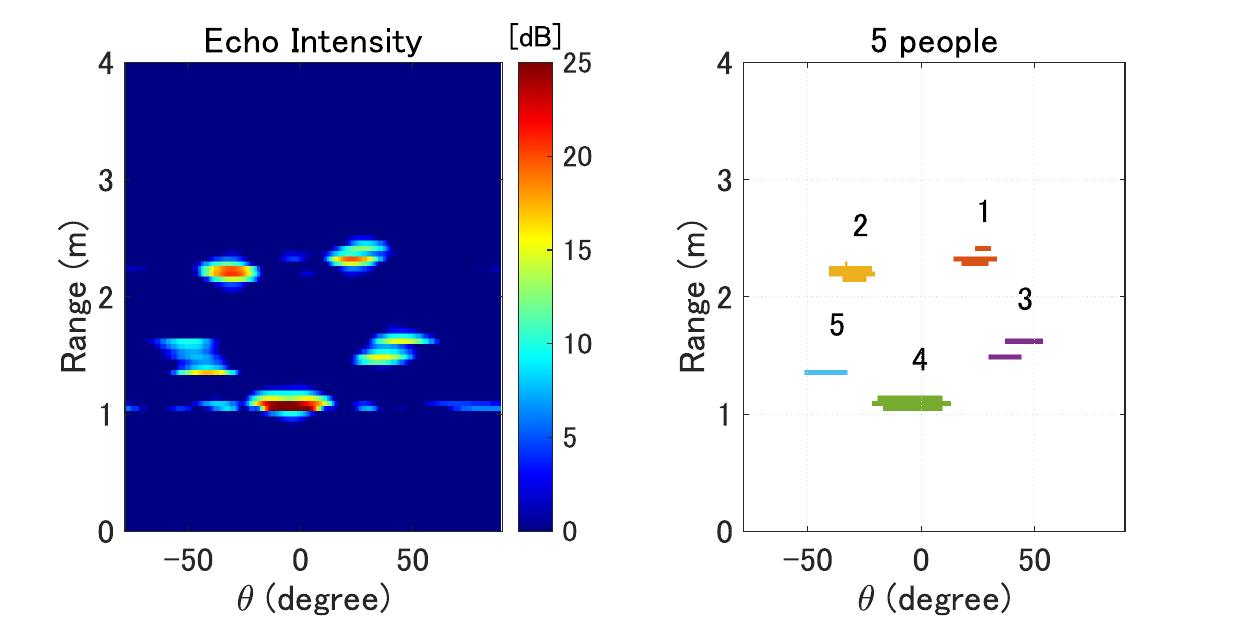}
          \\(c)\hspace{40mm}(d)\vspace{3mm}
      \end{minipage}
  \end{tabular}
  \caption{Experiment 5 with seven participants (a), their seating positions (b), the radar image (c), and clusters estimated using the proposed respiratory-space clustering (d).}
  \label{fig:radar1_case4}
\end{figure}  
\begin{figure}[t]
    \centering
    \begin{tabular}{c}
        \begin{minipage}[t]{\hsize}
            \centering
            \includegraphics[width=\textwidth]{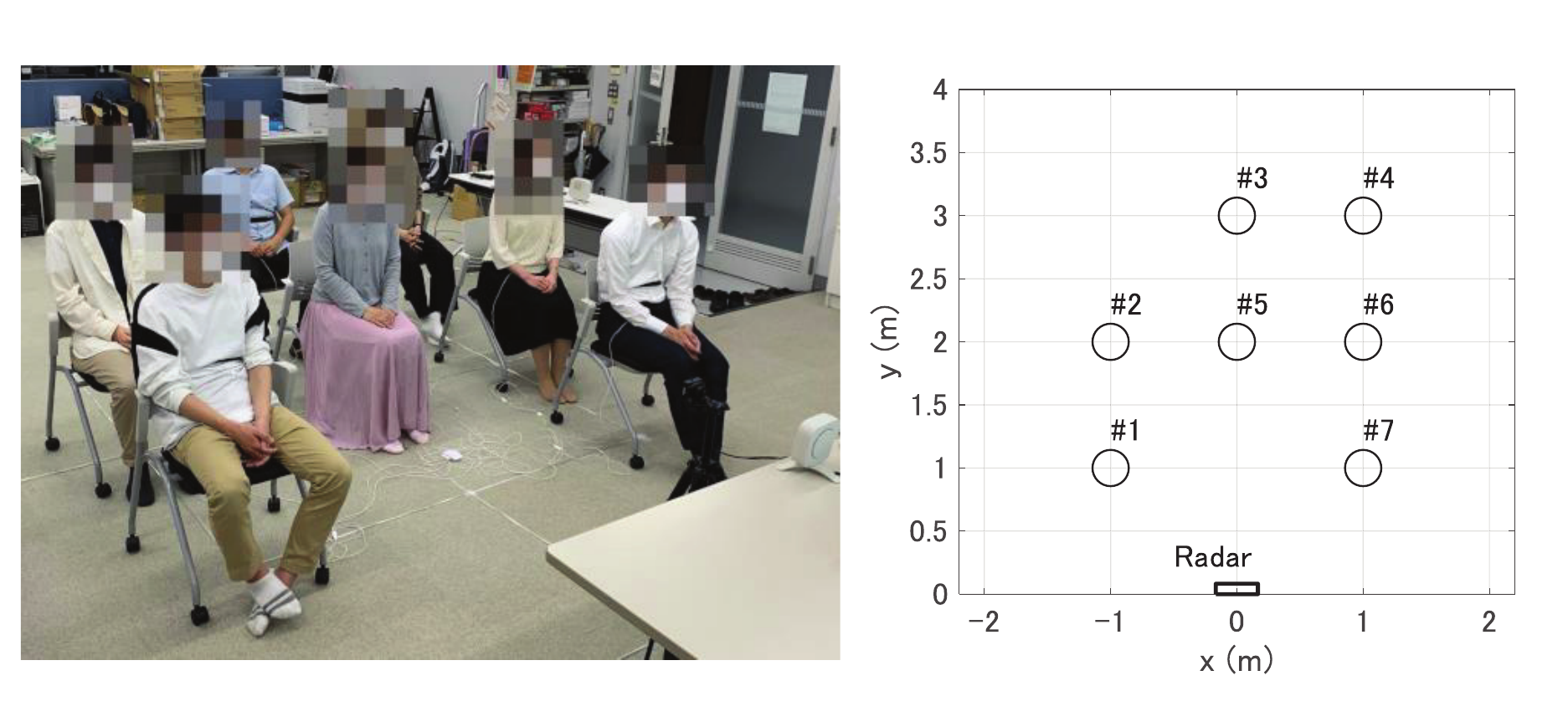}
            \\\hspace{5mm}(a)\hspace{40mm}(b)\vspace{3mm}
        \end{minipage}
    \end{tabular}
    \\
    \begin{tabular}{c}
      \begin{minipage}[t]{\hsize}
          \centering
          \includegraphics[width=\textwidth]{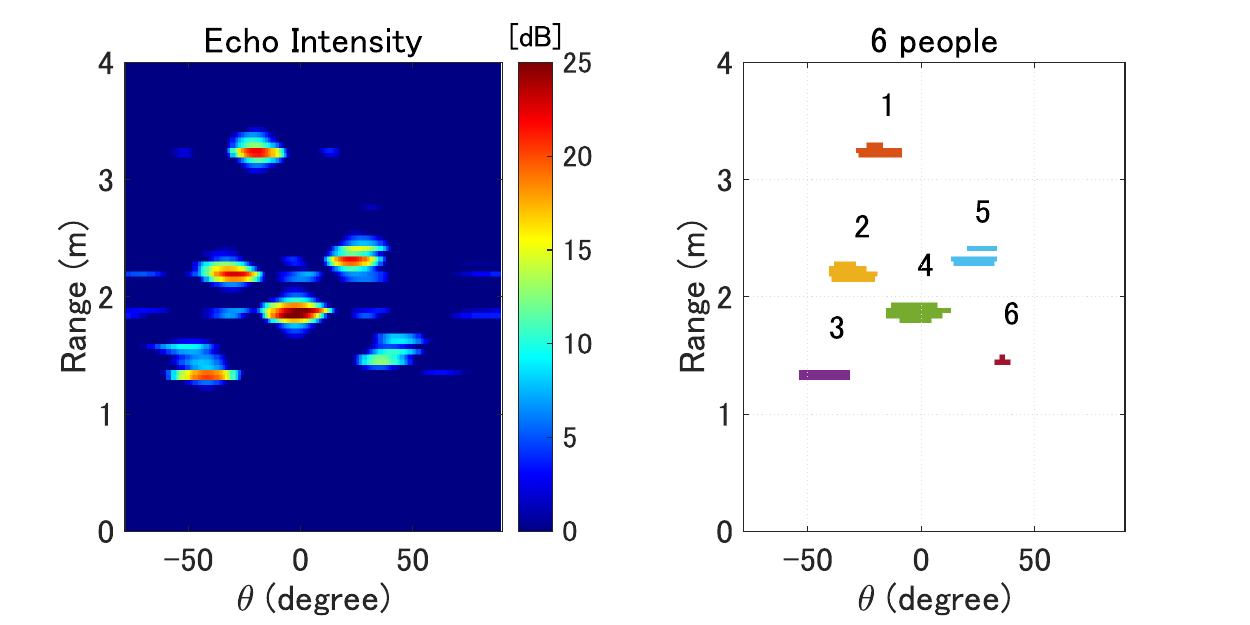}
          \\(c)\hspace{40mm}(d)\vspace{3mm}
      \end{minipage}
  \end{tabular}
  \caption{Experiment 6 with seven participants (a), their seating positions (b), the radar image (c), and clusters estimated using the proposed respiratory-space clustering (d).}
  \label{fig:radar1_case5}
\end{figure}  
\begin{figure}[t]
    \centering
    \begin{tabular}{c}
        \begin{minipage}[t]{\hsize}
            \centering
            \includegraphics[width=\textwidth]{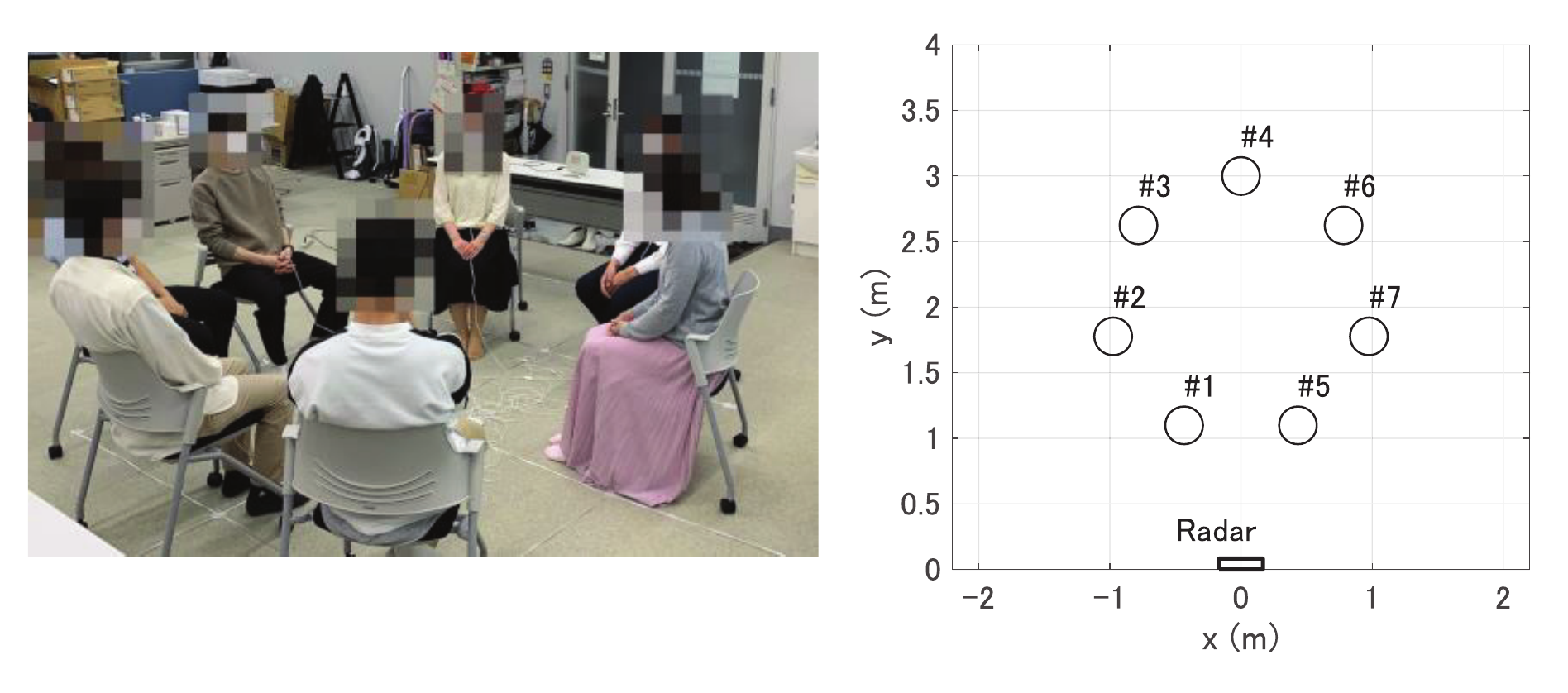}
            \\\hspace{5mm}(a)\hspace{40mm}(b)\vspace{3mm}
        \end{minipage}
    \end{tabular}
    \\
    \begin{tabular}{c}
      \begin{minipage}[t]{\hsize}
          \centering
          \includegraphics[width=\textwidth]{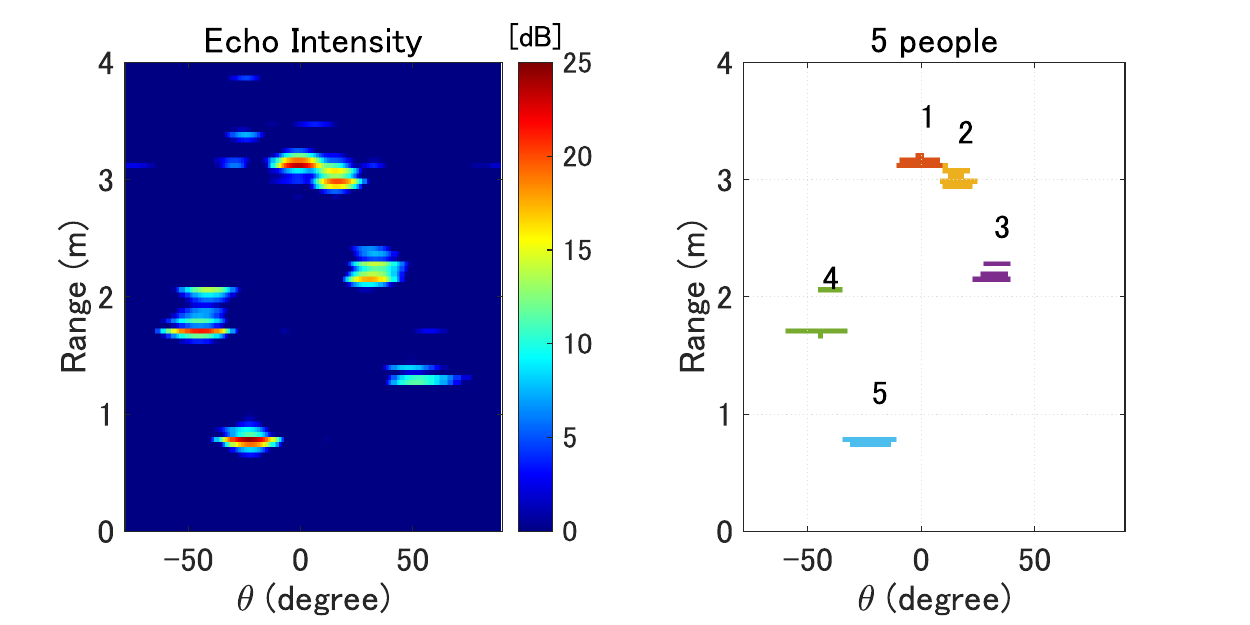}
          \\(c)\hspace{40mm}(d)\vspace{3mm}
      \end{minipage}
  \end{tabular}
  \caption{Experiment 7 with seven participants (a), their seating positions (b), the radar image (c), and clusters estimated using the proposed respiratory-space clustering (d).}
  \label{fig:radar1_case6}
\end{figure}

    To investigate the performance of the proposed method for other layouts of the participants, we conducted five additional measurements with different layouts, which are referred to as experiments 3, 4, 5, 6, and 7. Figures. \ref{fig:radar1_case2}-\ref{fig:radar1_case6} show the photograph of the measurement scenario in (a), the actual layout of participants in (b), the radar image in (c), and the clusters obtained using the proposed method (d) for each experiment. In these experiments, seven participants were instructed to breathe normally as in experiments 1 and 2. As seen in the figures, the number of people was estimated to be 7, 7, 5, 6, and 5 in experiments 3, 4, 5, 6, and 7, respectively. The number was correctly estimated in experiments 3 and 4, whereas the echoes from a few people were not detected in experiments 5, 6, and 7. These results are explained by shadowing; an echo from a person is blocked by the body of another person. Despite this limitation due to shadowing, the proposed method was demonstrated to be able to measure the physiological signals of multiple people.

  \subsection{Evaluation of the Respiratory Interval Estimation Accuracy}
  A commonly used indicator of the respiratory status is the respiratory rate, which is the number of breaths per minute and proportional to the reciprocal of the respiratory interval. We evaluated the accuracy of the proposed method in terms of instantaneous respiratory intervals that are not averaged over a period of 1 minute. We applied our proposed method to the radar signal in estimating the respiratory intervals, and evaluated the accuracy of the estimated respiratory intervals by comparing them with the values obtained using belt-type contact respirometers, which are assumed to be the true values.

\begin{figure}[t]
    \centering
    \includegraphics[width=0.49\textwidth]{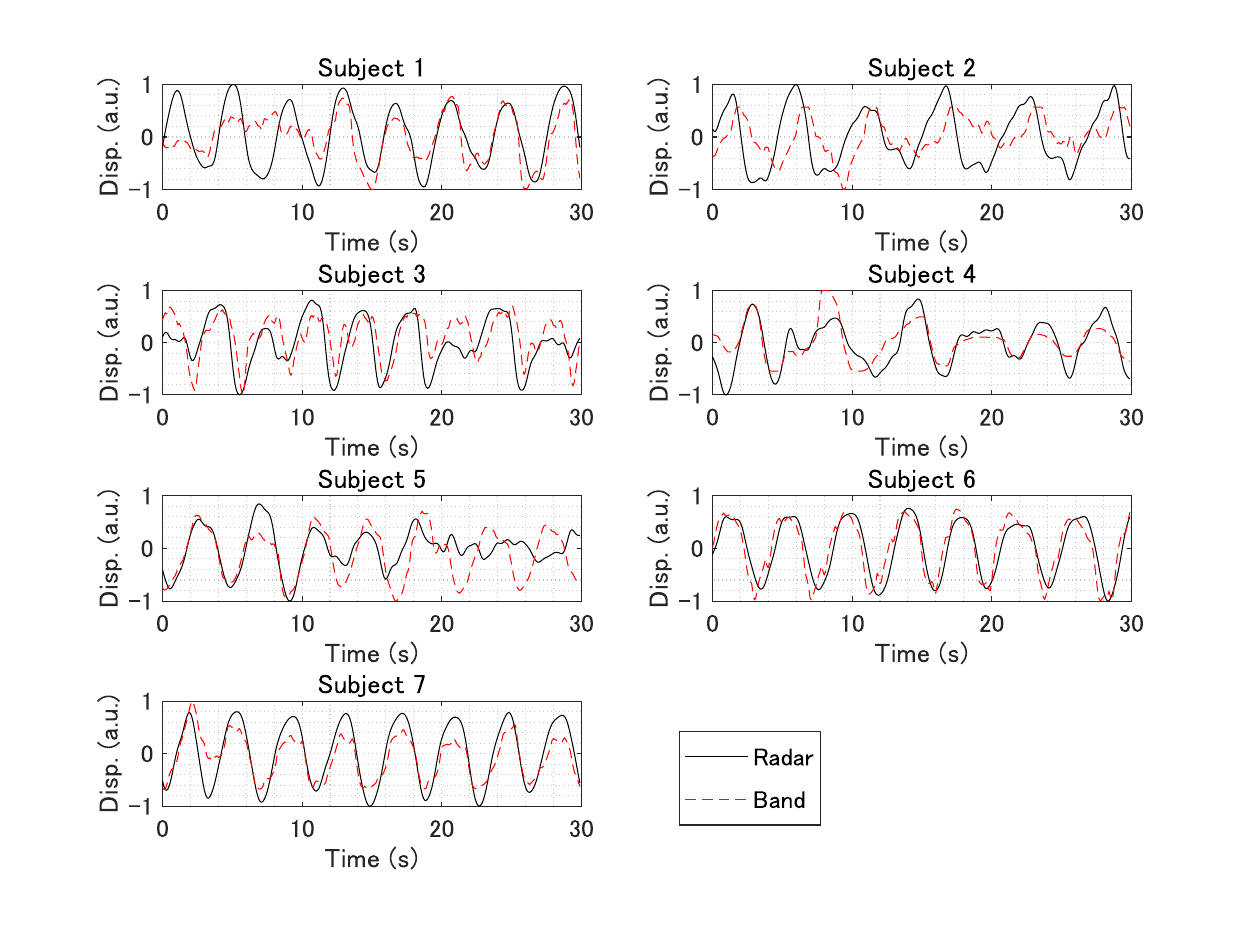}
    \caption{Normalized displacements of participants obtained using a radar system and belt-type contact respirometers (experiment 1).}
    \label{fig:displacement}
\end{figure}
  \begin{figure}[t]
    \centering
    \includegraphics[width=0.49\textwidth]{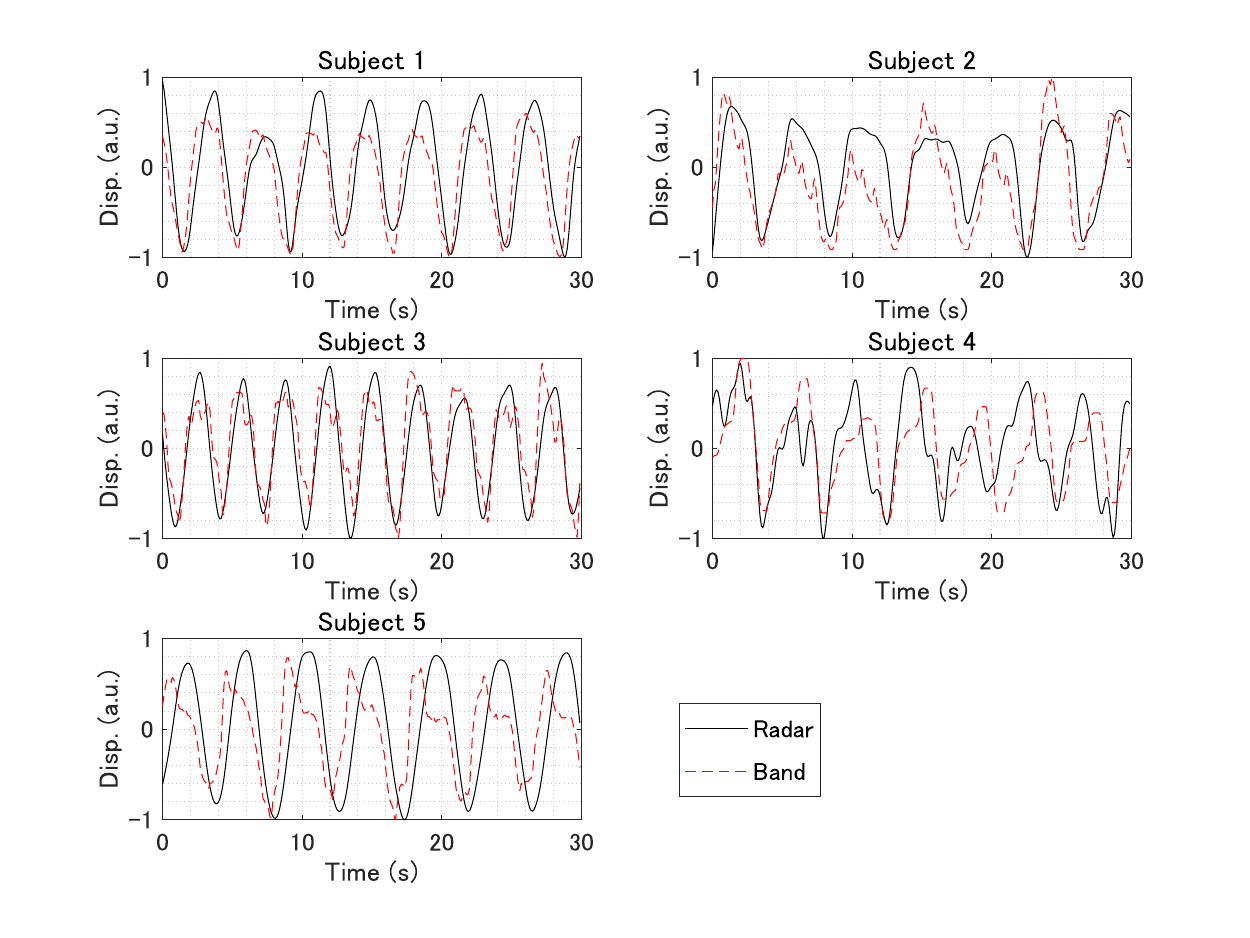}
    \caption{Normalized displacements of participants obtained using a radar system and belt-type contact respirometers (experiment 2).}
    \label{fig:displacement2}
  \end{figure}
  
  Figure \ref{fig:displacement} and \ref{fig:displacement2} present normalized displacements of the participants that are obtained using the radar system and respirometers in experiments 1 and 2. Note that the figures show the displacements only for 30 s that are extracted from the total measurement time. 
Table \ref{tab:jikkenn1234567} gives the actual number of participants $M_\mathrm{tr}$ and the number of participants estimated using the proposed method $M_\mathrm{est}$ for the seven experiments.
In addition, Table \ref{tab:jikkenn1234567} summarizes the root-mean-square error (RMSE) in estimating the respiratory intervals using the proposed method for each participant. Note that the RMSE is not presented in the table when the participant was not detected.
The average RMSE for all participants in the seven experiments was 196 ms, which is sufficiently smaller than a typical respiratory interval (e.g., 3--5 s), demonstrating that our system can estimate the respiratory interval with high accuracy. The results also indicate that the proposed method can estimate the respiratory intervals of the detected people accurately even when the number of people is incorrectly estimated.

  \begin{table}[t]
    \caption{Root-mean-square error in estimating the respiratory interval of each participant using the proposed method.}
    \begin{center}
        \scalebox{0.95}{
      \begin{tabular}{|c|c|c|ccccccc|}\hline
        Exp. &\multirow{2}{*}{$M_{\mathrm{tr}}$}&\multirow{2}{*}{$M_{\mathrm{est}}$}& \multicolumn{7}{c|}{RMSE for each participant (ms)}\\\cline{4-10}

        No. & & & 1 & 2 & 3 & 4 & 5 & 6 & 7\\

        \hline
                   1&7&7&121&112&51&541&306&34&428\\
        \hline
                   2&5&5&54&134&66&154&66&-&-\\
        \hline
                   3&7&7&168&141&115&169&73&226&226\\
        \hline
                   4&7&7&401&217&168&118&51&227&321\\
        \hline
                   5&7&5&268&131&-&141&-&144&115\\
        \hline
                   6&7&6&154&113&-&90&118&182&270\\
        \hline
                   7&7&5&-&124&197&453&799&-&240\\
        \hline
      \end{tabular}
        }
    \end{center}
    \label{tab:jikkenn1234567}
  \end{table}

  \subsection{Performance Evaluation of the Proposed Method for Different Radar Positions}
  \label{subsec:Experiment with multiple radar systems}
We investigate the effect of the radar position on the performance of the proposed method by analyzing two additional datasets with a radar system located at different positions $(x_2,y_2)$ (position 2) and $(x_3, y_3)$ (position 3), where $x_2=2$ m, $y_2=2$ m, $x_3=-2$ m, and $y_3=2$ m. The layout of the participants is the same as that in experiment 1 (Fig. \ref{fig:sokutei1}).
  \begin{figure}[t]
    \centering
    \begin{tabular}{c}
        \begin{minipage}[t]{\hsize}
            \centering
            \includegraphics[width=\textwidth]{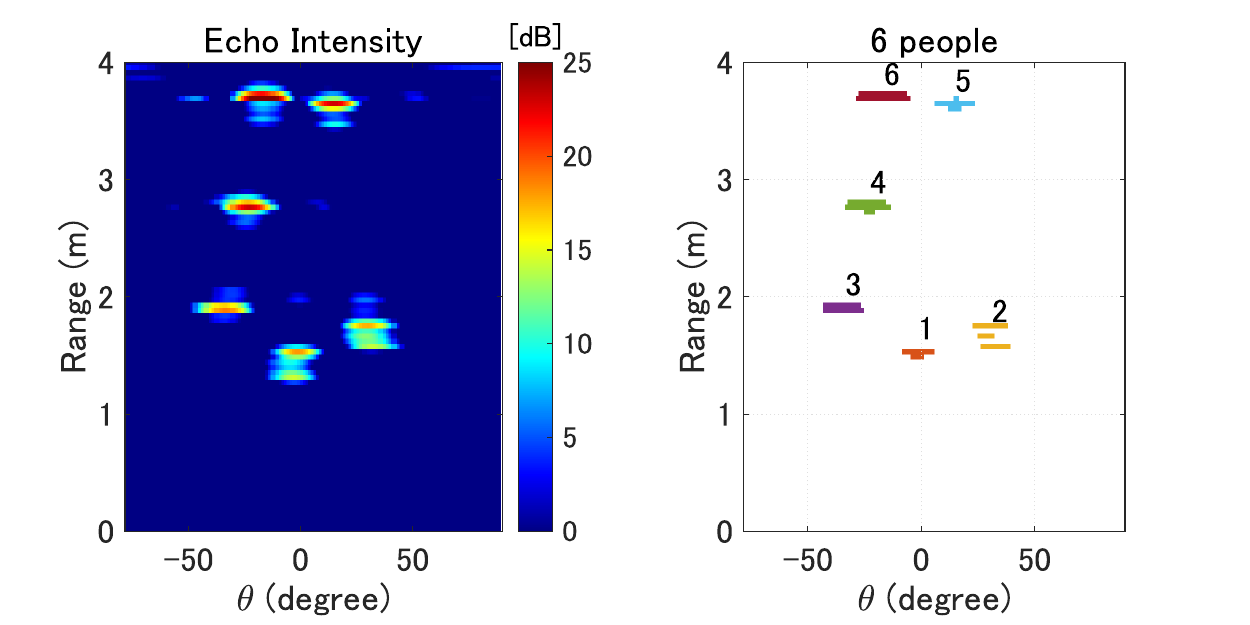}
            \\(a)\hspace{40mm}(b)\vspace{3mm}
        \end{minipage}
    \end{tabular}
    \\
    \begin{tabular}{c}
      \begin{minipage}[t]{\hsize}
          \centering
          \includegraphics[width=\textwidth]{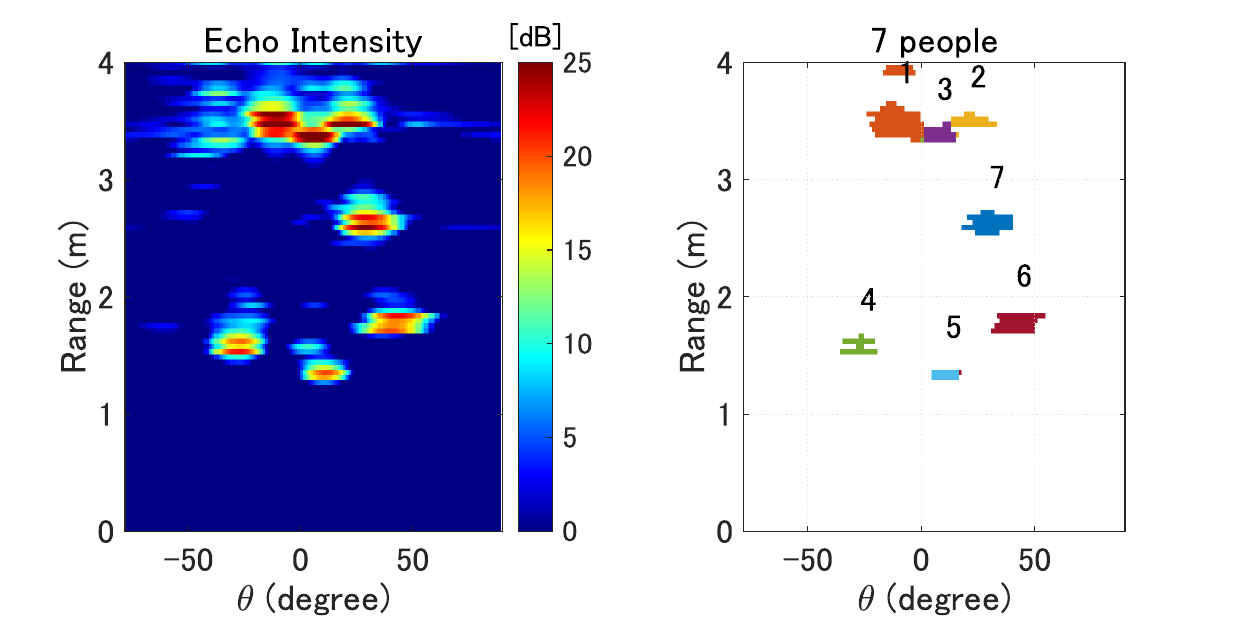}
          \\(c)\hspace{40mm}(d)\vspace{3mm}
      \end{minipage}
  \end{tabular}
  \caption{Radar images obtained using the radar system located at position 2 (a) and position 3 (c) and clusters estimated using the proposed method and radar system located at position 2 (b) and position 3 (d).}
  \label{fig:radar2andradar3}
\end{figure}  
Figure \ref{fig:radar2andradar3} shows the radar images in (a) and (c) and the clusters estimated using the proposed method in (b) and (d). Panels (a) and (b) are obtained with the radar system in position 2 whereas panels (c) and (d) are obtained with the radar system in position 3. When the radar was placed in position 3, all seven participants were correctly detected, whereas one of the participants was not detected when the radar was placed in position 2. This result can be explained by the shadowing as mentioned in Section IV-C. The body of a person who is located between the radar system and another person can block the echo, resulting in a missing cluster in the cluster image. This result indicates that although the proposed method works regardless of the radar position, the echo shadowing can prevent an echo from being detected.

  \subsection{Performance of the Proposed Method for Non-Static People}
  \begin{figure}[t]
    \centering
    \begin{tabular}{c}
        \begin{minipage}[t]{\hsize}
            \centering
            \includegraphics[width=\textwidth]{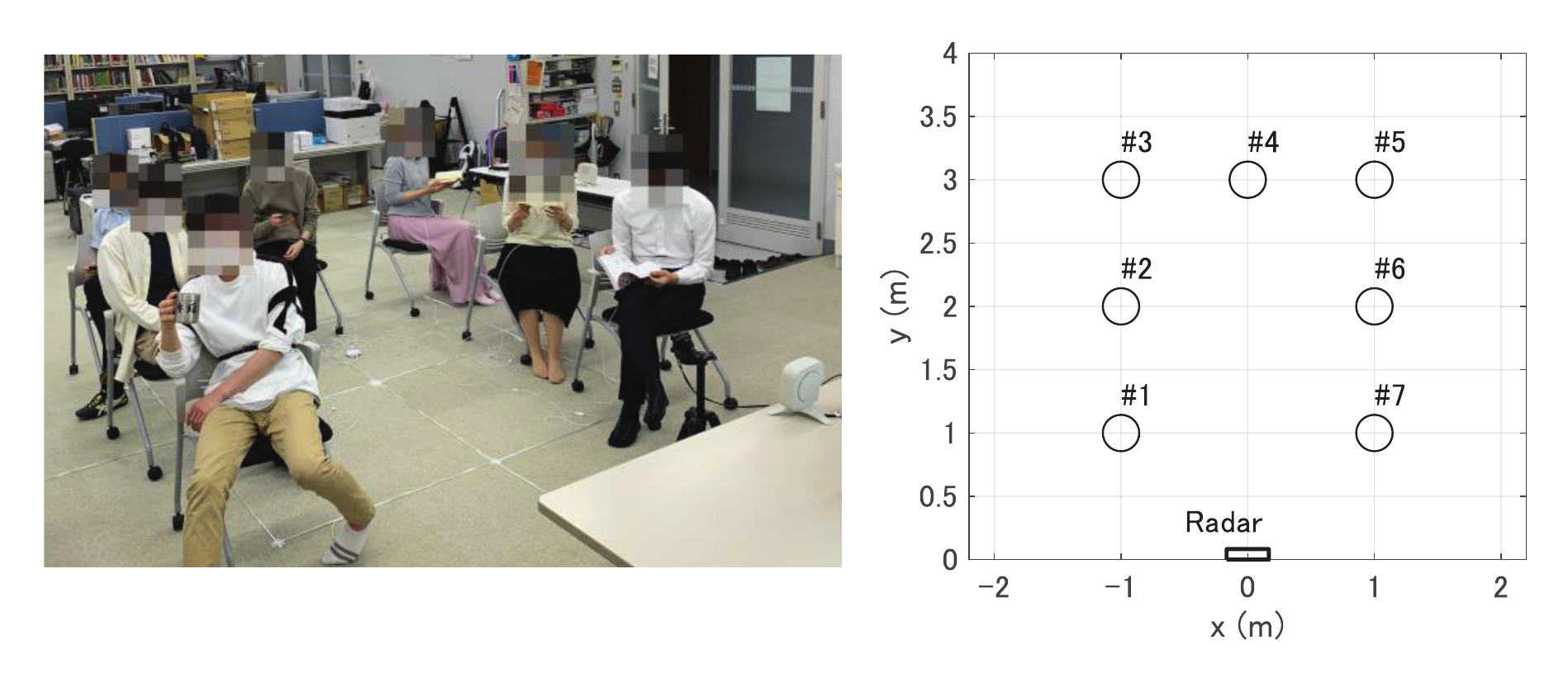}
            \\\hspace{5mm}(a)\hspace{40mm}(b)\vspace{3mm}
        \end{minipage}
    \end{tabular}
    \\
    \begin{tabular}{c}
      \begin{minipage}[t]{\hsize}
          \centering
          \includegraphics[width=\textwidth]{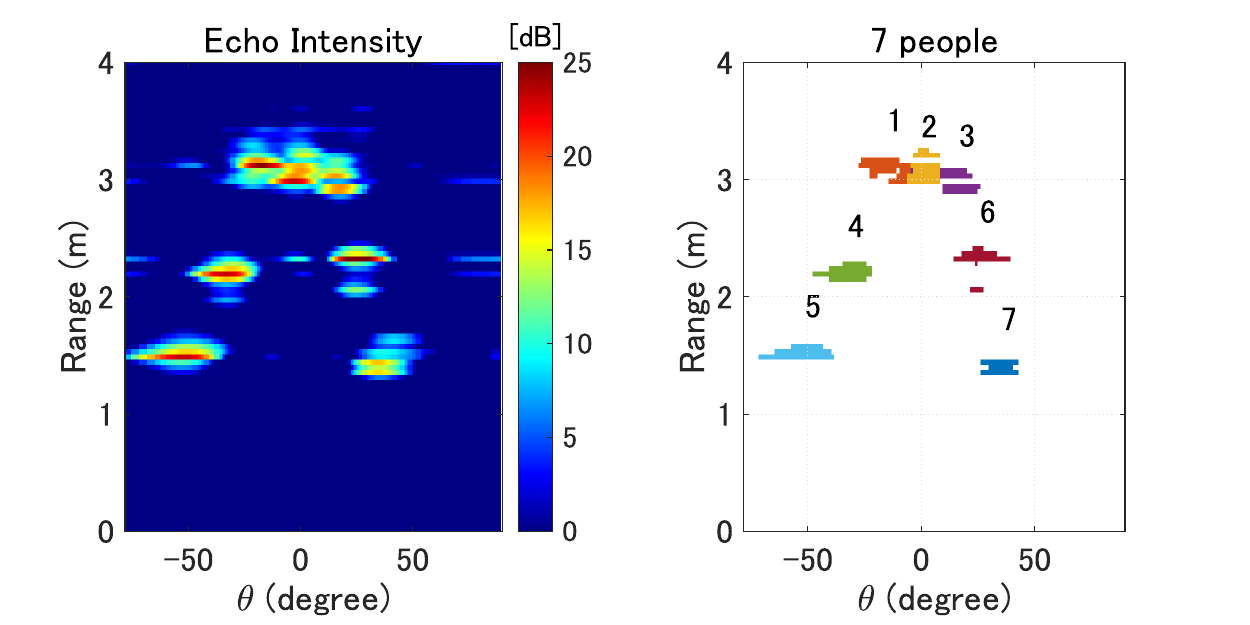}
          \\(c)\hspace{40mm}(d)\vspace{3mm}
      \end{minipage}
  \end{tabular}
  \caption{Experimental validation in a non-static situation where the participants perform small movements, such as the situations of holding a smartphone or reading a book (a), their actual positions (b), the radar image (c), and clusters estimated using the proposed respiratory-space clustering (d).}
  \label{fig:radar1_case0}
\end{figure}

So far, we have assumed a relatively static situation where the participants were instructed to remain still while breathing normally. In this subsection, we evaluate the performance of the proposed method when the participants were instructed to behave normally while drinking coffee, reading a book, and handling a smartphone. Panels (a) and (b) in Fig. \ref{fig:radar1_case0} respectively show the photograph of the measurement scenario and the actual layout of participants. Note that the layout of the participants is the same as that in experiment 1. Panels (c) and (d) in Fig. \ref{fig:radar1_case0} respectively show a radar image and the estimated clusters.
As seen in Fig.~\ref{fig:radar1_case0}, all seven participants were correctly separated and detected even in the non-static situation. 
This result indicates the possible applicability of the proposed method to a realistic scenario. Additional study is necessary to investigate the performance of the proposed method in realistic situations so that the proposed method can be extended to perform well in practical settings; this is left as future work.

  In this paper, we assumed that the radar measurements were made under a condition with little to no multipath propagation. In real life situations, however, the effect of multipath propagation must be taken into account. Some studies reported radar-based physiological measurement techniques that work even in a multipath-rich environment \cite{peerreview3,peerreview4,peerreview1,peerreview2}. 
  It is important to improve the accuracy of the proposed method when it is applied to measurement data recorded in a multipath-rich environment.

  \section{Conclusion}
  In this study, we developed a noncontact respiration monitoring system that simultaneously measures the respiration of several people, even when the number and locations of people are unknown. To evaluate the proposed respiration monitoring system, we measured the respiration of multiple human targets in eight experiments involving five or seven participants. The obtained results demonstrated that the use of the proposed clustering method in the 4D respiratory space was able to separate echoes accurately and estimate the number and locations of the participants unless some echoes were blocked through shadowing. Using the proposed clustering method instead of the conventional clustering method, the accuracy in estimating the number of people was improved by 88.5\% and 81.4\% in experiments 1 and 2, respectively, indicating that the proposed method is effective in estimating the number of participants. Belt-type respirometers were used in the experiments to evaluate the accuracy of the respiratory intervals estimated using the radar system and the proposed method. In the experiments, the proposed method had an average RMSE of 196 ms, which is sufficiently smaller than a typical respiratory interval (e.g., 3-5 s). The obtained results demonstrate that the clustering method and system proposed in this study are effective for measuring the respiration of several people accurately and simultaneously even when the number and locations of the people are unknown.

  \section*{Acknowledgment}
  A preprint of this manuscript has been posted on arXiv (arXiv:2101.12422 [eess.SP]).

    \section*{Ethics Declarations}
      This study was approved by the Ethics Committee of the Graduate School of Engineering, Kyoto University (permit no.~201916). Informed consent was obtained from all participants in the study.

  \newpage
    \begin{IEEEbiography}[{\includegraphics[width=1in,height=1.25in,clip,keepaspectratio]{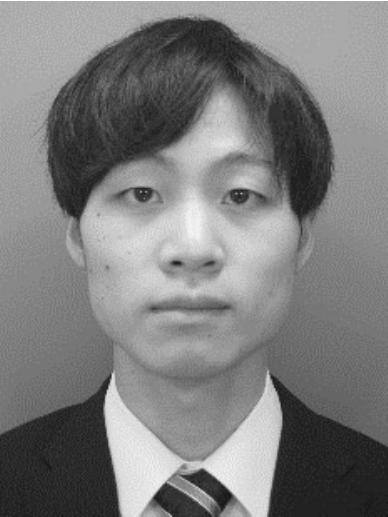}}]{Takato Koda}
      received a B.E.~degree in electrical and electronic engineering from Kyoto University, Kyoto, Japan, in 2020. He is currently working toward an M.E. degree in electrical engineering at the Graduate School of Engineering, Kyoto University.
    \end{IEEEbiography}
    
    \begin{IEEEbiography}[{\includegraphics[width=1in,height=1.25in,clip,keepaspectratio]{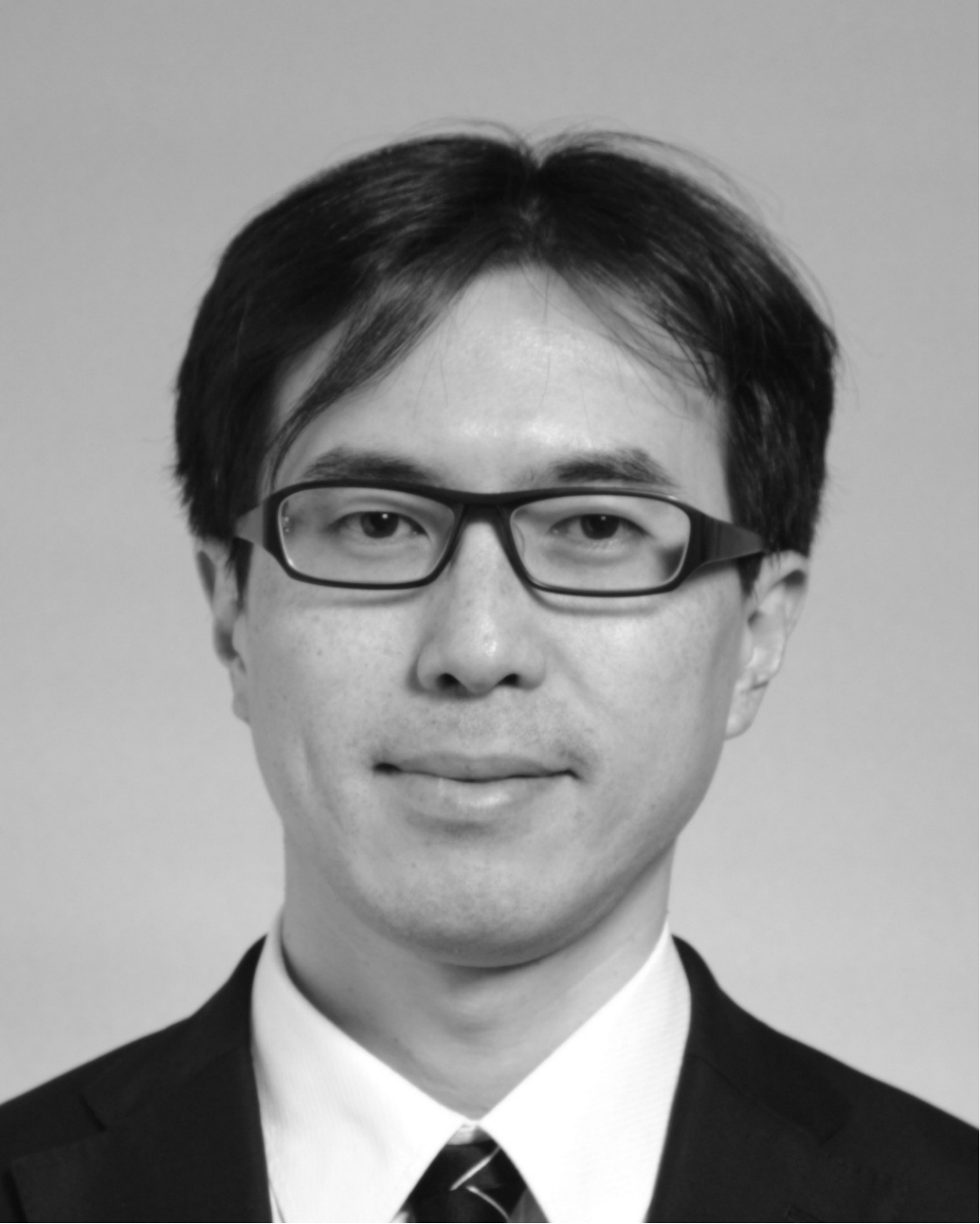}}]{Takuya Sakamoto} (Senior Member, IEEE) received a B.E.~degree in electrical and electronic engineering from Kyoto University, Kyoto, Japan, in 2000 and M.I.~and Ph.D.~degrees in communications and computer engineering from the Graduate School of Informatics, Kyoto University, in 2002 and 2005, respectively.
    
    From 2006 to 2015, he was an Assistant Professor at the Graduate School of Informatics, Kyoto University. From 2011 to 2013, he was also a Visiting Researcher at Delft University of Technology, Delft, the Netherlands. From 2015 to 2018, he was an Associate Professor at the Graduate School of Engineering, University of Hyogo, Himeji, Japan. In 2017, he was also a Visiting Scholar at the University of Hawaii in Manoa, Honolulu, HI, USA. From 2018, he has been a PRESTO Researcher at the Japan Science and Technology Agency, Kawaguchi, Japan. Currently, he is an Associate Professor at the Graduate School of Engineering, Kyoto University. His current research interests are system theory, inverse problems, radar signal processing, radar imaging, and the wireless sensing of vital signs.
    
    Dr. Sakamoto was a recipient of the Best Paper Award from the International Symposium on Antennas and Propagation (ISAP) in 2012 and the Masao Horiba Award in 2016. In 2017, he was invited as a semi-plenary speaker to the European Conference on Antennas and Propagation (EuCAP) in Paris, France.
    \end{IEEEbiography}

    \begin{IEEEbiography}[{\includegraphics[width=1in,height=1.25in,clip,keepaspectratio]{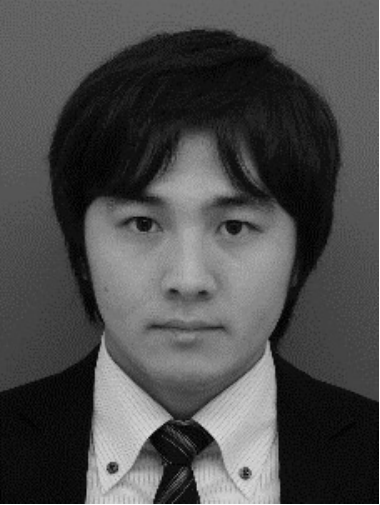}}]{Shigeaki Okumura}
        received his B.E. degree inelectrical engineering from Kyoto University, Kyoto, Japan, in 2013 and M.I. and Ph.D. degrees incommunications and computer engineering from theGraduate School of Informatics, Kyoto University, in 2015 and 2018, respectively. He has been with MaRI Co., Ltd. since 2019. His research interests includeradar and audio signal processing and the noncontactmeasurement of vital signs.
      \end{IEEEbiography}
      
      \begin{IEEEbiography}[{\includegraphics[width=1in,height=1.25in,clip,keepaspectratio]{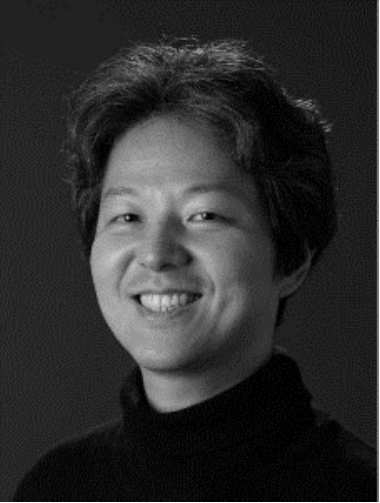}}]{Hirofumi Taki}
         (Member, IEEE) received the M.D. and Ph.D. degrees in informatics from Kyoto University, Japan, in 2000 and 2007, respectively. He was an Assistant Professor with the Graduate School of Informatics, Kyoto University, and an Associate Professor with the Graduate School of Biomedical Engineering, Tohoku University. He founded MaRI Co., Ltd. in 2017, and has been the CEO ever since. His research interest includes digital signal processing in measurement of biological information.
      \end{IEEEbiography}
      \EOD
    
     \end{document}